\pdfoutput=1
\newcommand*{\ATLASLATEXPATH}{}
\documentclass[cernpreprint, texlive=2016, USenglish]{\ATLASLATEXPATH atlasdoc}
 
\usepackage[subfigure=true]{\ATLASLATEXPATH atlaspackage}
\usepackage{\ATLASLATEXPATH atlasbiblatex}
 
\usepackage{\ATLASLATEXPATH atlasphysics}
\usepackage{pdflscape,amssymb}
\usepackage{rotating}
 
\graphicspath{{logos/}{figures/}}
 
\usepackage{multirow}
\usepackage{ANA-STDM-2017-20-PAPER-defs}

\AtlasTitle{Search for electroweak diboson production in association with a high-mass dijet system in semileptonic final states
in $pp$ collisions at $\sqrt{s}=13$~\TeV\ with the ATLAS detector}

\AtlasAbstract{
This paper reports on a search for electroweak diboson ($WW/WZ/ZZ$) production in association with a high-mass dijet system,
using data from proton--proton collisions at a center-of-mass energy of $\sqrt{s}=13$~\TeV.
The data, corresponding to an integrated luminosity of \lumi, were recorded with the ATLAS detector in 2015 and 2016 at the Large Hadron Collider.
The search is performed in final states in which one boson decays leptonically, and the other boson decays hadronically.
The hadronically decaying $W/Z$ boson is reconstructed as either two small-radius jets or one large-radius jet
using jet substructure techniques.
The electroweak production of $WW/WZ/ZZ$ in association with two jets is measured with an observed~(expected) significance of
\sigfVBSobs~(\sigfVBSexp) standard deviations,
and the fiducial cross section is measured to be
$\fidVBSobsNom \pm \fidVBSobsStat (\mathrm{stat.}) ^{+\fidVBSobsSystUp}_{-\fidVBSobsSystDn} (\mathrm{syst.})$~fb.
}
 
\author{The ATLAS Collaboration}
 
\AtlasRefCode{STDM-2017-20}
 
\PreprintIdNumber{CERN-EP-2019-072}

\AtlasJournalRef{\PRD 100, 032007 (2019)}
\AtlasDOI{10.1103/PhysRevD.100.032007}

\hypersetup{pdftitle={ATLAS document},pdfauthor={The ATLAS Collaboration}}
 
\begin{document}
\LEcontact{Richard Keeler rkeeler@uvic.ca}
\maketitle
 
\tableofcontents

\section{Introduction}
\label{sec:intro}
% The next lines are included from the .//./sections/Intro.tex input file
Vector-boson scattering~(VBS) is a key process for probing the non-Abelian gauge structure of
the electroweak~(EW) sector of the Standard Model~(SM),
since it involves both the self-couplings of the vector bosons and their coupling with the Higgs boson.
In the absence of the SM Higgs boson, the amplitudes for VBS would
increase as a function of partonic center-of-mass energy and ultimately violate unitarity~\cite{Lee:1977yc,Lee:1977eg}.
The discovery of a Higgs boson in 2012 at the LHC~\cite{HIGG-2012-27,CMS-HIG-12-028},
with measured properties~\cite{HIGG-2014-06,HIGG-2013-17,CMS-HIG-14-009,CMS-HIG-14-018}
consistent with those of the SM Higgs boson,
represents a major milestone in the understanding of electroweak symmetry breaking.
The study of the VBS process provides an important check of the SM by testing whether the Higgs mechanism is the sole source of electroweak symmetry breaking.
Theories of new phenomena beyond the SM that alter the quartic gauge couplings~\cite{Eboli:2003nq,Eboli:2006wa},
or include the presence of additional resonances~\cite{Chang:2013aya,Espriu:2012ih},
predict enhancements of VBS  at high transverse momentum
of the vector bosons and at high invariant mass of the diboson system.

\begin{figure}[!h]
\centering
\subfigure[]{\includegraphics[width=0.325\textwidth]{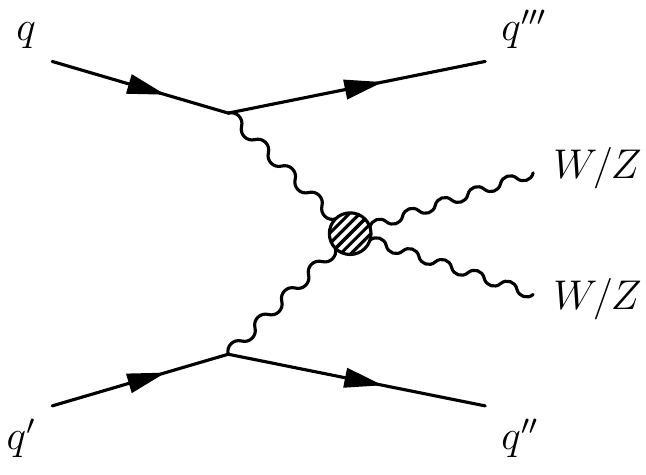}}
\subfigure[]{\includegraphics[width=0.325\textwidth]{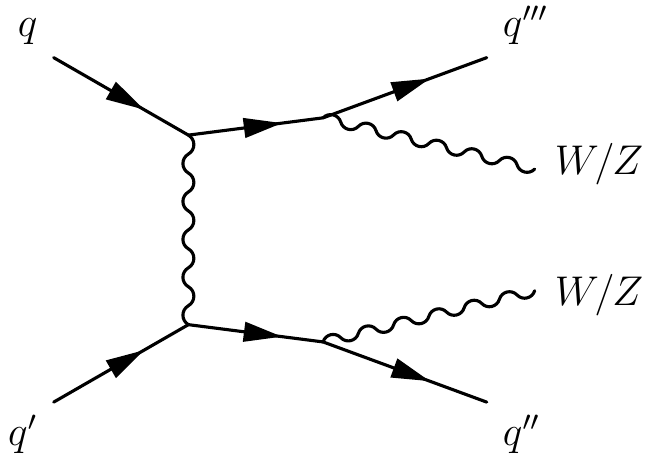}}
\subfigure[]{\includegraphics[width=0.325\textwidth]{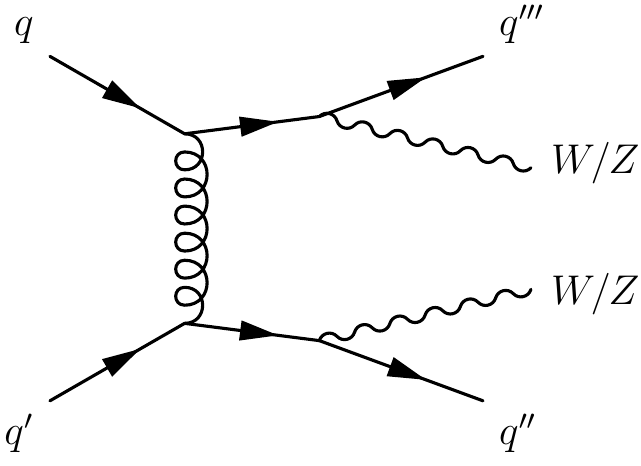}}
\caption{\label{fig:diag} Representative Feynman diagrams for (a) EW \VVjj\ production via VBS, (b) EW \VVjj\ production
via non-VBS contribution, and (c) QCD \VVjj\ production.
}
\end{figure}

The experimental signature of VBS is characterized by the presence of a pair of vector
bosons and two forward jets, \VVjj~($V=W, Z, \gamma$), with a large
separation in rapidity of jets and a large dijet invariant mass.
Multiple processes can produce the same final state of two bosons and two jets.
The production of $VVjj$ at tree level has an EW contribution involving only
electroweak-interaction vertices,
and a strong contribution (QCD-induced) involving two strong-interaction vertices.
The EW production is further divided into two components.
The first component is EW VBS production with actual scattering of the two electroweak bosons.
The scattering occurs via quartic gauge vertices,
or triple gauge vertices involving the $s$- or $t$-channel exchange of a Higgs boson or a $W/Z$ boson.
The second component is  EW non-VBS production that has electroweak vertices only, but
where the two bosons do not scatter.
The EW non-VBS component cannot be separated from the EW VBS component in a gauge invariant way~\cite{Barger:1990py}
and contributes significantly to the total cross section.
It is therefore included in the signal generation.
Representative Feynman diagrams at tree level are shown in Figure~\ref{fig:diag}.
Both the ATLAS and CMS Collaborations have searched  for experimental evidence of VBS.
So far, electroweak \VVjj\ production is only  observed in the same-sign $W^\pm W^\pm jj$
channel~\cite{CMS-SMP-17-004} and $WZjj$ channel~\cite{STDM-2017-23}
in the fully leptonic final states using data collected at a center-of-mass energy of $\sqrt{s} = 13$~\TeV.
Evidence of electroweak \VVjj\ production is also obtained in the
$W^\pm W^\pm jj$~\cite{STDM-2013-06,STDM-2014-11,CMS-SMP-13-015} and
$Z\gamma jj$~\cite{CMS-SMP-14-018} channels using $pp$ collisions at  $\sqrt{s} = 8$~\TeV.
Limits on fiducial cross sections of electroweak \VVjj\ production are reported for the $WZjj$~\cite{CMS-SMP-18-001,STDM-2014-02},
$ZZjj$~\cite{CMS-SMP-17-006}, $Z\gamma jj$~\cite{STDM-2015-21} and $W\gamma jj$~\cite{CMS-SMP-14-011} channels.
Constraints on anomalous quartic gauge couplings are reported in
Refs.~\cite{CMS-SMP-13-015,STDM-2013-06,STDM-2014-11,CMS-SMP-13-015,STDM-2014-02,CMS-FSQ-13-008,
STDM-2015-09,CMS-SMP-14-018,CMS-SMP-14-011,STDM-2015-21,CMS-SMP-18-006}.
 
Reference~\cite{STDM-2015-09} reports a study similar to the one in this paper, albeit focused on EW production of
\VVjj\ in the \WVlvqq\ channel only and performed at $\sqrt{s} = 8$~\TeV.
This paper presents a study of the EW production of \VVjj~($V=W,Z$) with the vector-boson pair
decaying semileptonically.
A larger data sample is used and additional diboson signal processes with similar final states are included.

Three $VV$ semileptonic decay channels are explored:
a $Z$ boson decaying into a pair of neutrinos, $Z\to\nu\nu$;\footnote{To simplify the notation, antiparticles are not explicitly labeled in this paper.}
a $W$ boson decaying into a charged lepton~(an electron or muon, denoted by $\ell$)
and a neutrino, $W\to\ell\nu$;
and a $Z$ boson decaying into a pair of light charged leptons, $Z\to\ell\ell$.
In all cases, the other vector boson $V$ is required to decay into a pair of quarks, \Vtqq,
leading to \ZVvvqq, \WVlvqq\ and \ZVllqq\ final states.
These processes overlap in the fiducial region of the measurement
because of the geometrical acceptance of the detector for leptons and jets.
The decay  channels are selected as
0-, 1- and 2-lepton final states, where
the \olep~(\tlep) final state receives only contribution from \WVlvqq~(\ZVllqq) processes,
and the \zlep\ final state receives about equal contributions from \WVlvqq\ and \ZVvvqq processes.

Two different reconstruction techniques for the \Vtqq decay are considered: resolved and merged.
The resolved reconstruction attempts to identify two separate small-radius jets~(small-$R$ jet denoted by $j$) of hadrons
from the \Vtqq decay, while the merged reconstruction uses jet substructure techniques to identify the
\Vtqq decay reconstructed as a large-radius jet~(large-$R$ jet denoted by $J$).
The latter applies when the momentum transfer in \VVjj\ production is high, and as a consequence
the $qq$ pair from the $V$ boson decay is collimated.
In this case, hadrons from the two quarks overlap in the detector and are more efficiently
reconstructed as a single large-$R$ jet.
In total, six final states are included in this study: 0-, 1- and 2-lepton final states, each using
resolved or merged \Vtqq\ reconstruction techniques.

In order to extract the signal and to measure the cross section for the EW production of \VVjj\ in a fiducial volume,
multivariate discriminants, which combine
observables sensitive to the kinematics of the VBS process, are used to separate EW-induced \VVjj\ production from
QCD-induced \VVjj\ production.
 
This analysis measures the cross section of EW \VVjj\ production in a region of kinematic phase space
close to the acceptance of the detector.
Fiducial cross sections are measured in the 0-, 1- and 2-lepton channels,
where lepton refers to $e$ and $\mu$.
Final states with $V$ decaying into one or more $\tau$-leptons (both leptonically and hadronically decaying $\tau$-leptons)
are included as signal, but the contribution of $V$ from top quark decay is not considered as signal.
% End of text imported from the .//./sections/Intro.tex input file
 
\section{ATLAS detector}
\label{sec:detector}
% The next lines are included from the .//./sections/ATLAS.tex input file
\newcommand{\AtlasCoordFootnote}{
The ATLAS experiment uses a right-handed coordinate system with its origin at the nominal interaction point (IP)
in the center of the detector and the $z$-axis along the axis of the beam pipe.
The $x$-axis points from the IP to the center of the LHC ring,
and the $y$-axis points upwards.
Cylindrical coordinates $(r,\phi)$ are used in the transverse plane,
$\phi$ being the azimuthal angle around the $z$-axis.
The pseudorapidity is defined in terms of the polar angle $\theta$ as $\eta = -\ln \tan(\theta/2)$.
Angular distance is measured in units of $\Delta R \equiv \sqrt{(\Delta\eta)^{2} + (\Delta\phi)^{2}}$.}

The ATLAS experiment is described in Ref.~\cite{PERF-2007-01}. ATLAS is a multipurpose detector with a forward–backward symmetric cylindrical geometry and a solid-angle\footnote{\AtlasCoordFootnote} coverage of nearly 4$\pi$.
The inner tracking detector~(ID), covering the region $|\eta| <$ 2.5, consists of a silicon pixel detector, a silicon microstrip detector and a straw-tube transition-radiation tracker.
The inner detector is surrounded by a thin superconducting solenoid providing a $2\,\text{T}$ magnetic field, and by a finely segmented lead/liquid-argon (LAr) electromagnetic calorimeter covering the region $|\eta| <$ 3.2.
A steel/scintillator-tile hadronic calorimeter provides coverage in the central region $|\eta| <$ 1.7.
The end-cap and forward regions are instrumented with LAr calorimeters
for both EM and hadronic energy measurements up to \(|\eta| = 4.9\).
A muon spectrometer~(MS) system incorporating large superconducting toroidal air-core magnets surrounds the calorimeters. Three layers of precision wire chambers provide muon tracking in the range $|\eta| <$ 2.7, while dedicated fast chambers are used for triggering in the region $|\eta| <$ 2.4.
The trigger system is composed of two stages~\cite{TRIG-2016-01}.
The first stage, implemented with custom hardware, uses information from calorimeters and muon chambers to reduce the event rate to a maximum of 100~kHz. The second stage, called the high-level trigger, reduces the data acquisition rate to about 1~kHz on average. The high-level trigger is software-based and runs reconstruction algorithms similar to those used in the offline reconstruction.
 
% End of text imported from the .//./sections/ATLAS.tex input file
 
\section{Data and Monte Carlo simulation}
\label{sec:Samples}
% The next lines are included from the .//./sections/Samples.tex input file
\subsection{Data}
\label{sec:data}
The data were collected with the ATLAS detector in 2015 and 2016 from $pp$ collisions at a center-of-mass energy
of $\sqrt{s} = 13$~\TeV, corresponding to a total integrated luminosity of \lumi.
 
The recorded \tlep\ channel and \olep\ channel events were selected
with a mixture of  either multiple single-electron or single-muon triggers with
varying transverse energy $\et$ (electron) and transverse momentum $\pt$ (muon) thresholds, and quality and isolation requirements, that depended on the LHC running conditions.
The lowest $\et$ or $\pt$ requirement without trigger prescaling was 26~\GeV\ for both the electrons and muons.
Events for the \zlep\ channel were recorded with non-prescaled missing transverse momentum~(\met) triggers
where the \met\ threshold depended on the LHC running conditions.
The lowest threshold used is 110~\GeV.
The \met\ triggers used are fully efficient for events passing the selection described below.
The \met\ triggers are also used in the \olep\ channel to compensate for single-muon trigger inefficiency
due to the difference in acceptance between the muon tracking and triggering.

Events in this analysis have all detector systems operating normally.
Collision vertices are formed from tracks with $\pt>400\,\MeV$,
and the one with the highest $\sum\pt^2$ of its associated tracks is selected as the primary vertex.

\subsection{Signal and background simulation}
\label{sec:mc}
Monte Carlo (MC) simulation is used to model signal and background processes.
The simulated samples are used to optimize the event selection, to develop the
multivariate discriminant, and to estimate the irreducible background yields.

The EW \VVjj\ signal samples were generated using \MGMCatNLO~2.4.3~\cite{Alwall:2014hca}
with amplitudes of $\mathcal{O}(\aew^6\aqcd^0)$, where \aew (\aqcd) is the EW (strong) coupling constant.
Both the VBS amplitudes and non-VBS amplitudes of the \VVjj\ process with one  boson decaying hadronically and
the other leptonically were included, using factorized on-shell decays for the gauge bosons.
The \textsc{NNPDF30LO}~\cite{Ball:2012cx} PDF set was used.
The parton showers and hadronization were modeled with \textsc{Pythia}~8.186~\cite{Sjostrand:2007gs} using the A14 set
of tuned parameters (tune) for the underlying event~\cite{ATL-PHYS-PUB-2014-021}.

The main background sources are $Z$ and $W$ bosons produced in association with jets (\Zjets and \Wjets),
as well as significant contributions from top quark production (both \ttbar pair and single-top)
and  QCD-induced vector-boson pair production.
The \Zjets and \Wjets events were simulated using the \textsc{Sherpa} 2.2.1~\cite{Gleisberg:2008ta} event generator.
Matrix elements were calculated for up to two partons at NLO and up to four partons at LO using the
\textsc{Comix}~\cite{Gleisberg:2008fv} and \textsc{OpenLoops}~\cite{Cascioli:2011va} programs.
QCD-induced diboson processes with one of the bosons decaying hadronically and the other leptonically were simulated using
\textsc{Sherpa}~2.2.1. They were simulated for up to one additional parton at NLO and up to three additional partons at LO
using the \textsc{Comix} and \textsc{OpenLoops} programs.
There is no overlap between the QCD-induced diboson samples and the EW \VVjj\ signal samples, as the former
include diagrams of $\mathcal{O}(\aew^4\aqcd^2)$.
For \Zjets, \Wjets and diboson simulation, the matrix-element calculations were merged with the \textsc{Sherpa} parton shower
using the ME+PS@NLO prescription~\cite{Hoeche:2012yf}.
The \textsc{NNPDF30NNLO}~\cite{Ball:2014uwa} PDF set was used in conjunction with a dedicated parton-shower tuning developed by the \textsc{Sherpa} authors.
For the \Zjets and \Wjets samples, boson decays into all lepton flavors ($e,\mu,\tau$) are included.
For the generation of top quark pairs, the \textsc{Powheg-Box}~v2~\cite{Nason:2004rx,Frixione:2007nw,Alioli:2010xd}
event generator with the CT10~\cite{CT10} PDF set
in the matrix-element calculations was used. Electroweak $t$-channel, $s$-channel and $Wt$-channel single-top-quark events
were generated using the \textsc{Powheg-Box}~v1 event generator~\cite{Alioli:2009je,Frederix:2012dh,Re:2010bp}.
This event generator uses the four-flavor scheme for the NLO matrix-element calculations together with the fixed four-flavor
PDF set CT10f4~\cite{CT10}.  For all top quark processes, top quark spin correlations are preserved~(for the $t$-channel,
top quark decay is simulated using \textsc{MadSpin}~\cite{Artoisenet:2012st}).
The parton showers, fragmentation, and underlying event were simulated using \textsc{Pythia}~6.428~\cite{Sjostrand:2006za}
with the \textsc{CTEQ6L1}~\cite{1126-6708-2002-07-012} PDF set and
the corresponding Perugia 2012 tune (P2012)~\cite{Skands:2010ak}.
The top quark mass was set to 172.5~\GeV. The \textsc{EvtGen}~v1.2.0 program~\cite{EvtGen} was used to simulate the decay of
bottom and charm hadrons for the \textsc{Powheg-Box} samples.
 
All simulated processes are normalized using the  currently available state-of-the-art theoretical predictions for their
cross sections.
Cross sections are calculated with up to next-to-next-to-leading-order (NNLO) QCD corrections for \Zjets and \Wjets
production~\cite{zxsec_NNLO}. Cross sections for diboson production are calculated at NLO including LO contributions with
two additional partons~\cite{Gleisberg:2008ta,Hoche:2010kg}.
The \ttbar production cross section is calculated at NNLO in QCD, including resummation of next-to-next-to-leading
logarithmic (NNLL) soft-gluon terms~\cite{ttbarxsec1,ttbarxsec2}.
The single-top production cross sections are calculated to NLO in QCD~\cite{topschxsec},
including the soft-gluon resummation at NNLL~\cite{wtxsec} for the $Wt$ process.

MC events were processed with a detailed detector simulation~\cite{SOFT-2010-01} based on \textsc{Geant4}~\cite{Agostinelli:2002hh}.
Additional inelastic simulated $pp$ collisions generated with \textsc{Pythia}~8.186
using the A2 set of tuned parameters~\cite{ATL-PHYS-PUB-2012-003} and the \textsc{MSTW2008LO}~\cite{Martin:2009iq} PDF set
were overlaid in order to model both
the in- and out-of-time effects from additional $pp$ collisions in the same and neighboring bunch crossings~(pileup).
MC samples are reweighted to match the pileup conditions in the data.
All simulated events are processed using the same reconstruction algorithms as the data.
% End of text imported from the .//./sections/Samples.tex input file
 
\section{Object reconstruction}
\label{sec:Objects}
% The next lines are included from the .//./sections/Objects.tex input file
Electrons are identified as isolated energy clusters in the electromagnetic calorimeter matched to
ID tracks, and are required to have transverse energy $\et>7$~\GeV\ and pseudorapidity $|\eta|<2.47$.
A likelihood-based requirement~\cite{PERF-2017-01} is imposed to reduce the background from
non-prompt electrons or hadrons misidentified as electrons. Electrons are classified as either `loose', `medium' or `tight'
according to the likelihood-based identification criteria described in Ref.~\cite{PERF-2017-01}.
 
Muons are reconstructed by a combined fit to the ID and MS tracks, and are required to have $\pt>7$~\GeV\ and $|\eta|<2.5$. Muons must pass identification requirements, based on the number of hits in the ID and MS subsystems, and the significance of the difference $|q/p_{\mathrm {MS}} - q/p_{\mathrm {ID}}|$~\cite{PERF-2015-10}, where $q$ is the charge and $p_{\mathrm {MS}}\ (p_{\mathrm {ID}})$ is the momentum of the muon measured in the MS (ID). Similarly to electrons, muons are classified as either `loose', `medium' or `tight', following the criteria in Ref.~\cite{PERF-2015-10}.
 
All electrons and muons are required to be isolated by using selections based on the sum of the $\pt$ of tracks (excluding the track associated with the lepton) in a cone of $\pT$-dependent size around their directions. The isolation selection criteria are designed to maintain a constant efficiency of 99\% in the $\pt$--$\eta$ plane for reconstructed leptons from $Z\to\ell\ell$ decays.
Furthermore, leptons are required to have associated tracks satisfying $|d_0/\sigma_{d_0}|<5~(3)$ and $|z_0\times\sin\theta|<0.5$ mm for electrons~(muons), where $d_0$ is the transverse impact parameter relative to the beam line, $\sigma_{d_0}$ is its uncertainty, and $z_0$ is the distance between the longitudinal position of the track along the beam line at the point where $d_0$ is measured and the longitudinal position of the primary vertex.
 
Three types of jets are employed in the analysis.
Two of them are reconstructed from three-dimensional topological clusters of energy deposits
in the calorimeter~\cite{PERF-2014-07} (small-$R$ jets and large-$R$ jets),
and the third type from inner-detector tracks (track jets). All three use the anti-$k_t$ algorithm~\cite{antikt_algorithm,Cacciari:2011ma}
but with different values of the radius parameter $R$.
Small-$R$ jets and large-$R$ jets are reconstructed independently from the same energy depositions for a given event.
The treatment of the resulting overlap is discussed further below.

Small-$R$ jets are reconstructed with a radius parameter of $R=0.4$. Energy- and $\eta$-dependent correction factors derived from MC simulations are applied to correct jets back to the particle level~\cite{PERF-2016-04}.
Pileup effects are corrected using a jet area method~\cite{PERF-2014-03,jetarea}.
Jets are required to have $\pT>20$~\GeV\ for $|\eta|<2.5$ and $\pT>30$~\GeV\ for $2.5<|\eta|<4.5$.
A jet vertex tagger~\cite{PERF-2014-03} is applied to jets with $\pT<60$~\GeV\ and $|\eta|<2.4$ in order to select only jets from the hard interaction which are associated with the primary vertex, and to suppress jets from pileup interactions.
This tagger uses information about tracks associated with the primary vertex and pileup vertices.

Small-$R$ jets containing $b$-hadrons are identified using a multivariate algorithm ($b$-tagging)~\cite{PERF-2012-04} which uses information such as track impact-parameter significance and the position of explicitly reconstructed secondary decay vertices.
The chosen $b$-tagging algorithm has an efficiency of 70\% for $b$-quark jets in simulated $t\bar{t}$ events, with a light-flavor jet rejection factor of about 380 and a $c$-jet rejection factor of about 12~\cite{ATL-PHYS-PUB-2016-012}.

Large-$R$ jets are reconstructed with the radius parameter increased to $R=1.0$.
In order to mitigate the effects of pileup and soft radiation, the large-$R$ jets are trimmed~\cite{jet_trimming}.
Trimming takes the original constituents of the jet and reclusters them using the $k_t$ algorithm \cite{kt_algorithm} with a smaller radius parameter, $R_{\mathrm {subjet}}$, to produce a collection of subjets. These subjets are discarded if they carry less than a specific fraction ($f_{\mathrm{cut}}$) of the original jet \pT.
The trimming parameters were optimized for $W/Z$ boson tagging and are $R_{\mathrm {subjet}} =0.2 $ and $f_{\mathrm{cut}}=5\%$.
The large-$R$ jet four-momenta are recomputed from the remaining subjets, and the jet energies are calibrated to particle level using correction factors derived from MC simulations~\cite{PERF-2012-02}.
The mass of a large-$R$ jet~($m_J$) is computed using a combination of calorimeter and tracking information~\cite{ATLAS-CONF-2016-035}.
Large-$R$ jets are required to have $\pt > 200$~\GeV\ and $|\eta|<2.0$.

Track jets have a radius parameter of $R= 0.2$~\cite{ATL-PHYS-PUB-2014-013}. Inner-detector tracks originating from the primary vertex,
with $\pT > 0.5$~\GeV\ and selected by impact parameter requirements, are used in the track jet reconstruction.
Track jets are required to satisfy $\pT > 20$~\GeV\ and $|\eta| < 2.5$.
The number of track jets is  an input to the multivariate discriminant described later.

An overlap-removal procedure is applied to the selected leptons and jets in order to prevent double-counting.
The jet is removed if an electron and a small-$R$ jet are separated by
$\Delta R<$ 0.2; the electron is removed if the separation satisfies 0.2 $<\Delta R<$ 0.4.
The jet is removed if a muon and a small-$R$ jet are separated by
$\Delta R<$ 0.2 and if the jet has less than three tracks or the energy and momentum differences between
the muon and the jet are small; otherwise the muon is removed if the separation satisfies $\Delta R<$ 0.4.
In order to prevent double-counting of energy from an electron inside a large-$R$ jet, the large-$R$ jet is
removed if an electron and a large-$R$ jet are separated by $\Delta R < 1.0$.
No overlap removal is applied between large-$R$ jets or track jets and small-$R$ jets.

Boson tagging is applied to large-$R$ jets in order to select those consistent with $V\to qq$ decays.
A $\pT$-dependent requirement is applied to the jet substructure variable $\DTwoBetaOne$,
which is defined as a ratio of two-point to three-point energy correlation functions~\cite{Larkoski:2014gra,Larkoski:2015kga} that are based on the energies and pairwise angular separations of the particles within a jet. This variable is optimized to distinguish between jets originating from a single parton and those coming from the two-body decay of a heavy particle.
A detailed description of the method and its optimization can be found in Ref.~\cite{JETM-2018-03}.
Large-$R$ jets from $V\to qq$ decays are  required to have a  jet mass $m_J$  in a $\pT$-dependent window
centered around the expected value of the boson mass.
The configuration of the boson tagging algorithm is called a working point,
which is designed to provide constant efficiency independent of the large-$R$ jet $\pT$ for the signals studied.
Two working points are used, one with 50\% efficiency and the other one with 80\% efficiency,
with corresponding misidentification rates for jets from multijet production of $\sim 2$\% and $\sim 10$\%, respectively.
 
The missing transverse momentum vector, \metvec, is calculated as the negative vectorial sum of the transverse momenta of
calibrated electrons, muons, and small-$R$ jets where the calibration already includes corrections for pileup.
Large-$R$ jets and track jets are not included in the \metvec\ calculation in order to avoid double-counting of energy between the small-$R$ jets and large-$R$ jets. Energy depositions due to the underlying event and other types of
soft radiation are taken into account by constructing a `soft term' from ID tracks that are associated with the primary vertex but not used in any reconstructed object~\cite{PERF-2016-07}.
The track-based missing transverse momentum vector, $\mptvec$, is the negative vectorial sum of the transverse momenta of all good-quality inner-detector tracks that are associated with the primary vertex.
 
% End of text imported from the .//./sections/Objects.tex input file
 
\section{Event selection and background estimation}
\label{sec:SelectionBkg}
% The next lines are included from the .//./sections/Analysis.tex input file
Events are categorized into the 0-, 1- and 2-lepton channels depending on the number of selected electrons and muons.
In addition to a leptonically decaying candidate \Vlep, events in all three channels  are required to
contain a hadronically decaying candidate \Vhad, and two additional small-$R$ jets (referred to as tagging-jets).
The \Vhad candidate is reconstructed as either two small-$R$ jets~($V\to jj$) in a resolved selection,
or one large-$R$ jet~($V\to J$) in a merged selection, and those jets are referred to as \Vhad\ jets.
Event selection criteria are chosen to guarantee the statistical independence of the channels and to maximize
the sensitivity of the analysis.
This selection results in nine non-overlapping distinct signal regions (SR):
one for each of the three lepton channels and three types of \Vhad\ selections (resolved, and low- and high-purity merged).
 
The event selection for all channels and background estimations is summarized in Table~\ref{tbl:evselTable}.
Further details are given below.
\subsection{Event selection}
\label{sec:selection_SR}
\begin{sidewaystable}[phtb]
\small
\centering
\caption{Summary of the event selection in the 0-, 1- and 2-lepton channels.}
\label{tbl:evselTable}
\begin{tabular}{l | c | c | c }
\toprule
Selection & \zlep & \olep & \tlep \\
\midrule
\multirow{2}{*}{Trigger} & \met\ triggers & Single-electron triggers & Single-lepton triggers \\
&           & Single-muon or \met\ triggers & \\
\midrule
\multirow{2}{*}{Leptons} & 0 `loose' leptons    & 1 `tight' lepton with $\pT>27$~\GeV & 2 `loose' leptons with $\pT>20$~\GeV \\
&   with $\pT>7$~\GeV  & 0 `loose' leptons with $\pT>7$~\GeV & $\ge$ 1 lepton with $\pT>28$~\GeV \\
 
\midrule
\met\ & $>$ 200~\textrm{\GeV} & $>$ 80~\textrm{\GeV}  & --  \\
\midrule
\multirow{2}{*}{$m_{\ell\ell}$} & -- & -- & $83<m_{ee}<$ 99~\textrm{\GeV} \\
&  & & $(-0.0117\times \pT^{\mu\mu} + 85.63~\textrm{\GeV}) < m_{\mu\mu}< (0.0185\times \pT^{\mu\mu}+94~\textrm{\GeV})$  \\
\midrule
Small-$R$ jets &  \multicolumn{3}{c}{$\pt > 20$~\GeV\ if $|\eta| < 2.5$, and $\pt > 30$~\GeV\ if $2.5 < |\eta| < 4.5$} \\
\midrule
Large-$R$ jets &  \multicolumn{3}{c}{$\pt> 200$ \textrm{\GeV}, $|\eta|<2$  }  \\
\midrule
$\Vhad \to J$   & \multicolumn{3}{c}{ $V$ boson tagging, \minwzJ}\\
$\Vhad \to jj$  & \multicolumn{3}{c}{$64 < m_{j j} <$ 106~\GeV, $jj$ pair with \minwzjj, leading jet with $\pT>40$~\GeV}\\
\midrule
\multirow{2}{*}{Tagging-jets} &  \multicolumn{3}{c}{$j\notin \Vhad$, not $b$-tagged, $\Delta R(J,j)>1.4$ }\\
&  \multicolumn{3}{c}{$\etajo \cdot \etajt <0$,  $\mjjtag>400$~\GeV, $\pT>30$~\GeV }\\
\midrule
Num. of $b$-jets & -- & 0 & -- \\
\midrule
\multirow{4}{*}{Multijet removal} & $\mpt > 50$~\GeV & \multirow{4}{*}{--} & \multirow{4}{*}{--} \\
& $\dPhiEtmissPtmiss<\pi/2$  &  &  \\
& $\min[\dPhiEtmissj]>\pi/6$ &  &  \\
& $\dPhiEtmissVhad>\pi/9$    &  &  \\
\bottomrule
\end{tabular}
\end{sidewaystable}
 
Signal events in the \zlep channel are typical of a hadronically decaying $V$ boson recoiling against
a large amount of missing transverse momentum stemming from either a \Ztvv\ decay or a \Wtlv\ decay,
where the lepton is outside the acceptance of the detector.
An initial selection is made by requiring $\met >200$~\GeV,
and rejecting events with electrons or muons passing the `loose' quality requirements.
The multijet background originates primarily from the presence of mismeasured jets
and non-collision phenomena. It is suppressed using a requirement on the value of the track-based
missing transverse momentum, \mpt\ $>50$~\GeV.
Three further  angular selection criteria are:
the azimuthal separation between the \metvec\ and \mptvec\ directions satisfies $\dPhiEtmissPtmiss<\pi/2$;
the azimuthal separation between the directions of \metvec\ and
the nearest small-$R$ jet satisfies $\min[\dPhiEtmissj]>\pi/6$; and
the azimuthal separation between the directions of \metvec\ and
the reconstructed hadronically decaying candidate \Vhad\ satisfies $\dPhiEtmissVhad>\pi/9$.
The multijet background is found to be negligible after these selections.

The \olep channel is typical of a leptonically decaying $W$ boson.
The \Wtlv\ candidates are selected by requiring one isolated
lepton~(electron or muon) satisfying the `tight' criteria with $\pT> 27$~\GeV.
Events are required to have \met$>80$~\GeV, and must not have any additional `loose' leptons.
In order to reconstruct the invariant mass of the $WV$ system, needed later to construct the multivariate discriminant,
the neutrino momentum four-vector is reconstructed by imposing a $W$ boson mass constraint on
the lepton--neutrino system. The neutrino transverse momentum components are set equal to the
missing transverse momentum of the event and the unknown $z$-component of the momentum ($p_z$) is obtained from the
resulting quadratic equation. The $p_z$ is chosen as either the smaller, in absolute value, of the two real solutions or,
if the solution is complex, its real part.

In the \tlep channel, the \Ztll\ candidates are identified by requiring two isolated same-flavor
leptons satisfying the `loose' criteria. The leading~(subleading) lepton must satisfy $\pT > 28\, (20)$~\GeV.
Opposite charges are required for the muon pairs but not for the electron pairs, since electrons are more susceptible
to charge misidentification due to the conversion of photons from bremsstrahlung, especially at high \pT.
The dilepton invariant mass is required to be consistent with that of the $Z$ boson: $83< m_{ee}<99$~\GeV\ in the case of electrons and
$(-0.0117\times \pT^{\mu\mu} + 85.63~\GeV) < m_{\mu\mu}< (0.0185\times \pT^{\mu\mu}+94~\GeV)$ in the case of muons.
The $\pT$-dependent requirement on $m_{\mu\mu}$ recovers the selection efficiency at high $\pT^{\mu\mu}$, which would otherwise fall due to the degraded dimuon
invariant mass resolution~\cite{EXOT-2016-29}.

The merged selection is applied as the first step in identifying a \Vhad candidate. If  an event is not selected, then the resolved selection is used.
The order is motivated by a smaller background expectation in the merged analysis.
Selecting  the jets that form a \Vhad candidate first and then selecting the tagging-jets from
the pool of remaining jets results in an analysis with a higher sensitivity compared with doing the selection
in the reverse order. The \Vhad\ candidates are selected in three different non-overlapping channels.

Merged selection events are required to have at least one large-$R$ jet.
Next the boson tagging discussed in Section~\ref{sec:Objects} is applied to select the $V\to qq$ decays.
Two SRs are defined, one for events passing the 50\% working point of the boson tagging requirement
and the other for events failing the 50\%, but passing the 80\% working point requirement.
The former is called the high-purity~(HP) signal region, and the latter the low-purity~(LP) signal region.
Given the different but overlappping $W$ and $Z$ boson tagging requirements, large-$R$ jets are required to satisfy
either $W$ or $Z$ boson tagging.
If multiple \Vhad\ candidates are selected, the one minimizing \minwzJ\ is selected.

The resolved selection events are required to have two small-$R$ signal jets
with a dijet invariant mass lying in the $m_{W /Z}$ window: $64 < m_{jj} < 106$~\GeV.
If multiple \Vhad\ candidates are selected, the one minimizing \minwzjj\ is used.
At least one of the jets forming the  selected \Vhad\ candidate must have $\pT >40$~\GeV,
in order to improve the separation between the signal and the background;
otherwise the event is not selected.

After  selecting  the \Vhad\ candidate, tagging-jets are selected from the remaining small-$R$ jets
that fail the $b$-tagging described in Section~\ref{sec:Objects}.
For the merged selection, all small-$R$ jets with $\Delta R(J, j) < 1.4$ are excluded before the tagging-jets selection.
Tagging-jets are required to be in opposite hemispheres, $\etajo \cdot \etajt<0$, and
the invariant mass  of the two tagging-jets must satisfy $\mjjtag > 400$~\GeV.
If there is more than one pair of jets satisfying these requirements, the one with the highest \mjjtag\ value is chosen.
In order to suppress the contribution from pileup interactions, both tagging-jets
from the selected pair must have $\pT >30$~\GeV; otherwise the event is rejected,

Finally, \olep\ channel events are rejected if any of the small-$R$ jets
in the event is identified as a $b$-jet prior to the \Vhad\ candidate and tagging-jets selection.
This reduces the contributions from top quark production.
\subsection{Data control regions and background estimation}
\label{sec:bkgCR}
The dominant backgrounds for the \olep channel are  \Wjets and $t\bar t$ production;
for the \tlep channel it is \Zjets production;
while in the \zlep channel, they all contribute significantly.
Smaller background contributions for the \olep channel arise from multijet background. Single-top and QCD-induced diboson production
is a small background for all three lepton channels.
The background contributions are estimated using a combination of MC and data-driven techniques.
The shapes of kinematic variable distributions are taken from MC simulations in all cases except for the multijet background in the \olep channel.

A $Z$+jets control region~(ZCR)  is defined for each of the three SRs in the \tlep\ channel
by reversing the $m_J$ or $m_{jj}$ requirement.
Events in each of the CRs are selected in exactly the same way as those in their corresponding SRs
except for the requirement on $m_J$ or $m_{jj}$.
For the merged selection, the leading large-$R$ jet mass is required to be
outside the large-$R$ jet mass window of the 80\% working point of the $W/Z$ boson tagging.
For the resolved selection, a requirement of $50<m_{jj}<64$~\GeV\ or $m_{jj}>106$~\GeV\ is applied.
These CRs are dominated by the $Z$+jets contribution, with a purity higher than $95$\% in all regions.
They are therefore used to constrain its contribution in signal regions through simultaneous fits as discussed in Section~\ref{sec:results}.

Three $W$+jets control regions~(WCRs) are formed from events satisfying the \olep\ signal region selection except for the
invariant mass requirement of the \Vhad\ candidate, similar to the ZCRs.
Approximately 86\% and 77\% of the selected events are from $W$+jets production in the merged and resolved categories of the \olep channel,
respectively.
The remaining events are primarily from \ttbar production.

The three \ttbar control regions~(TopCRs) consist of events satisfying the signal region selection of the \olep channel
except for the $b$-jet requirement, which is inverted.
These CRs are dominated by \ttbar production, with a purity of 79\% and 59\% for merged and resolved categories respectively,
and the remainder are from single-top, $V$+jets or diboson production, for both the merged and the resolved event topologies.

In the \zlep channel, it is not possible to define pure control regions for \Wjets, \Zjets and \ttbar processes,
thus events falling into the mass sideband regions of the \Vhad, similar to WCRs and ZCRs,
form three different CRs~(referred to as VjjCR), one for each of the corresponding SRs.

The contribution from multijet production primarily consists of events with jets or photon conversions misidentified as
leptons or real but non-prompt leptons from decays of heavy-flavor hadrons.
This contribution is negligible in all regions, except for the resolved \olep\ SR.
The fake-factor background method of Ref.~\cite{EXOT-2016-28} is used to estimate the multijet background contribution
in the resolved topology of the \olep channel.
The estimated multijet contribution is about 10\% of the total background in the resolved \olep\ SR.

The \mjjtag\ spectra of simulated \Wjets (\Zjets) events are not well modeled by the MC simulation in the
WCRs~(ZCRs) for the three \Vhad\ selections in the 1-lepton (2-lepton) channel.
A data-driven procedure is applied to the simulated \Wjets and \Zjets events to correct for this shape mismodeling.
Reweighting factors are derived from WCRs and ZCRs as a function of \mjjtag,
and applied to all SRs and CRs (for 0-, 1-, and 2-lepton regions) in the MC simulation of \Wjets and \Zjets events, respectively.
The non-\Wjets(\Zjets) contributions are subtracted from the spectra in data.
Then the reweighting factors as a function of \mjjtag\ are determined by performing a linear fit
to the ratios of data to simulation in the control regions.
The reweighting is done separately for the merged and resolved analyses.
For \Wjets, the reweighting factor ranges from 1.016~(1.024) at $\mjjtag = 400$~\GeV\ to 0.47~(0.53)
at $\mjjtag = 3000$~\GeV\ in the resolved~(merged) analysis.
For \Zjets, the reweighting factor ranges from 1.071~(1.062) at $\mjjtag = 400$~\GeV\ to 0.42~(0.36)
at $\mjjtag = 3000$~\GeV\ in the resolved~(merged) analysis.

Additional reweighting factors are needed for the MC simulation of \Wjets and \Zjets events in the \zlep\
channel because the phase space is so different between the 0-lepton selection and the 1- and 2-lepton selections
that the reweightings described above are not applicable.
These additional reweightings are derived from MC simulation
as the ratio of the numbers of \Wjets (\Zjets) events in the \olep (\tlep) and \zlep channels,
and are applied to the MC simulation of \Wjets (\Zjets) events in the \zlep\ channel.
Good agreement between the prediction from MC simulation
and the data in the VjjCR is achieved only after the two reweightings have been applied.
Unless stated otherwise, the final reweighted \Wjets and \Zjets simulated events are used everywhere in the analysis.
% End of text imported from the .//./sections/Analysis.tex input file
 
\section{Multivariate analysis}
\label{sec:MVA}
% The next lines are included from the .//./sections/MVA.tex input file
A multivariate method is used  to enhance the separation between the signal and background.
The analysis uses the Toolkit for Multivariate Data Analysis, TMVA ~\cite{Hocker:2007ht}, and its implementation of the Boosted Decision Trees (BDTs) method.
BDTs are constructed, trained and evaluated in each lepton channel and analysis region separately.
The BDT training is carried out using simulated signal and all background MC samples.
However, the events in high-purity SR and low-purity SR are merged together for the BDT training due to an
insufficient number of MC events.
In order to make use of the complete set of simulated MC events for the BDT training and evaluation in an unbiased way,
the MC events are split for training and validation into two subsamples of equal size following the procedure in Ref.~\cite{HIGG-2016-29}.
The output distributions of the BDTs trained on the two subsamples are averaged for both the simulated and data events.
 
The input variables used for the BDTs are chosen in order to maximize the separation between signal and background,
and are summarized in Table~\ref{tbl:MVAinputs_boosted} and Table~\ref{tbl:MVAinputs_resolved},
for the merged and resolved category, respectively.
The distributions of input variables of the BDTs are compared between data and simulation,
and in general are found to be in good agreement.
The small-$R$ jets are labeled in decreasing \pT as `$j_1$' and `$j_2$' for the jets used to reconstruct
the hadronically decaying boson,  and as `tag, $j_1$' and `tag, $j_2$' for the tagging-jets.
The invariant mass and transverse momentum of the reconstructed $VV$ (\VVjj) system are denoted by \mVV~(\mVVjj)
and \ptVV~(\ptVVjj), respectively.
Angular variables are also considered, such as the pseudorapidity gap between the tagging-jets~(\detajjtag)
and between the small-$R$ \Vhad\ jets~(\detajj), the angular separation of the lepton and neutrino from the $W$ boson decay
(\dRlv) in the \olep\ channel, and the azimuthal angle between the directions of \metvec\ and the large-$R$ jet
(\dPhiEtmissJ) in the merged category of the \zlep\ channel.
A topological variable named boson centrality is also used, and it is defined as
$\zetaV = \mathrm{min}(\Delta \eta_{-}, \Delta \eta_{+})$, where
$\Delta \eta_{-}=\mathrm{min}[\eta(\Vhad),\eta(\Vlep)] - \mathrm{min}[\etajo,\etajt]$ and
$\Delta \eta_{+}=\mathrm{max}[\etajo,\etajt] - \mathrm{max}[\eta(\Vhad),\eta(\Vlep)]$.
The variable \zetaV\ has large values when the tagging-jets have a large separation in $\eta$, and when the
two boson candidates lie between the tagging-jets in $\eta$.
Variables sensitive to the quark--gluon jet separation are also included, such as the width
of the small-$R$ jets~($w$)~\cite{ATLAS-CONF-2016-034},
and the number of tracks associated with the jets~(\nTrksj).
The number of track jets, \nTrkJets, and the number of additional small-$R$ jets other than the \Vhad\ jets and tagging-jets,
\nJetsExtr, are also found to be useful for the BDTs.
In the \olep\ channel, the pseudorapidity of the lepton~(\etal) is also considered.

% The next lines are included from the .//sections/table_bdt_inputs.tex input file
\begin{table}[tb!]
\begin{center}
\caption{Variables used for the BDT discriminant in the merged analysis category of each lepton channel.}
\label{tbl:MVAinputs_boosted}
\begin{tabular}{l|ccc}
\toprule
Variable                                      &  \zlep   & \olep    & \tlep    \\
\midrule
 
\mjjtag      & \checkmark & --  & \checkmark \\
\\[-1em]
\detajjtag   & -- & --  &  \checkmark \\
\\[-1em]
\pttagjt     & \checkmark & \checkmark  & \checkmark \\
\\[-1em]
\mJ          & \checkmark & --  & -- \\
\\[-1em]
\DTwoBetaOne & \checkmark & --  & \checkmark \\
\\[-1em]
\MET         & \checkmark & --  & -- \\
\\[-1em]
\dPhiEtmissJ & \checkmark & --  &  -- \\
\\[-1em]
\etal        & -- & \checkmark & -- \\
\\[-1em]
\nTrkJets    & \checkmark & --  & -- \\
\\[-1em]
\zetaV       & -- & \checkmark  & \checkmark \\
\\[-1em]
\mVV         & -- & -- & \checkmark \\
\\[-1em]
\ptVV        & -- & -- &  \checkmark \\
\\[-1em]
\mVVjj       & -- & \checkmark &-- \\
\\[-1em]
\ptVVjj      & -- & -- &  \checkmark \\
\\[-1em]
\wttagjo     & \checkmark & -- & -- \\
\\[-1em]
\wttagjt     & \checkmark & -- & -- \\
 
\bottomrule
\end{tabular}
\end{center}
\end{table}

\begin{table}[tb!]
\begin{center}
\caption{Variables used for the BDT discriminant in the resolved analysis category of each lepton channel analysis.}
\label{tbl:MVAinputs_resolved}
\begin{tabular}{l|ccc}
\toprule
Variable                                      &  \zlep   & \olep    & \tlep    \\
\midrule
 
\mjjtag      & \checkmark & --  &  \checkmark \\
\\[-1em]
\detajjtag   & -- & --  &  \checkmark \\
\\[-1em]
\pttagjo     & \checkmark & \checkmark  & -- \\
\\[-1em]
\pttagjt     & \checkmark & \checkmark  & \checkmark \\
\\[-1em]
\detajj      &\checkmark  & \checkmark &  \checkmark \\
\\[-1em]
\ptjo        & \checkmark & --  & -- \\
\\[-1em]
\ptjt        & \checkmark & \checkmark  & \checkmark \\
\\[-1em]
\wtjo        & \checkmark & \checkmark  & \checkmark \\
\\[-1em]
\wtjt        & \checkmark & \checkmark  & \checkmark \\
\\[-1em]
\nTrksjo     & -- & \checkmark  & \checkmark \\
\\[-1em]
\nTrksjt     & -- & \checkmark  & \checkmark \\
\\[-1em]
\wttagjo     & \checkmark & \checkmark  & \checkmark \\
\\[-1em]
\wttagjt     & \checkmark & \checkmark & \checkmark \\
\\[-1em]
\nTrkstagjo  & -- & \checkmark  & \checkmark \\
\\[-1em]
\nTrkstagjt  & -- & \checkmark  & \checkmark \\
\\[-1em]
\nTrkJets    & \checkmark & -- & \checkmark \\
\\[-1em]
\nJetsExtr   & \checkmark & -- & -- \\
\\[-1em]
\MET         & \checkmark & --  & -- \\
\\[-1em]
\etal        & -- & \checkmark & -- \\
\\[-1em]
\dRlv        & -- & \checkmark  &  -- \\
\\[-1em]
\zetaV       & -- & \checkmark  & \checkmark \\
\\[-1em]
\mVV         & -- & -- & \checkmark \\
\\[-1em]
\mVVjj       & -- & \checkmark &-- \\
 
\bottomrule
\end{tabular}
\end{center}
\end{table}
% End of text imported from the .//sections/table_bdt_inputs.tex input file
 
% End of text imported from the .//./sections/MVA.tex input file
 
\clearpage
\section{Fiducial cross-section definition}
\label{sec:FidDef}
% The next lines are included from the .//./sections/FidXsec.tex input file
The fiducial phase space of the measurement is defined using stable final-state particles~\cite{ATL-PHYS-PUB-2015-013}.
Leptons produced in the decay of a hadron or its descendants are not considered
in the charged lepton requirement of the fiducial phase space. The fiducial selection is summarized in Table~\ref{tbl:vbs_fid_sel}
and details are given below.

Charged leptons are required to satisfy $\pt > 7$~\GeV\ and $|\eta| < 2.5$.
Jets are clustered from all final-state particles except prompt leptons, prompt neutrinos, and prompt photons using
the anti-$k_t$ algorithm.
Small-$R$ jets are required to have $\pT>20$~\GeV\ for $|\eta|<2.5$ and $\pT>30$~\GeV\ for $2.5<|\eta|<4.5$.
Jets within $\Delta R =0.2$ of any charged lepton
(as defined above) are rejected.
Jets containing a $b$-hadron, identified using `truth' information from the MC event record, are labeled as $b$-jets.
Large-$R$ jets are required to have $\pt > 200$~\GeV\ and $|\eta|<2.0$, and the same trimming algorithm as for the reconstruction-level large-$R$ jets is applied. No $\DTwoBetaOne$ requirement is applied to large-$R$ jets.
 
The selection of hadronically decaying bosons and tagging-jets follows the same steps and apply the same criteria as for reconstruction level,
as shown in Table~\ref{tbl:vbs_fid_sel}.

For the 0-, 1- and 2-lepton channels, the number of selected fiducial leptons is required to be 0, 1 and 2, respectively.
Events with additional leptons for the 1- and 2-lepton channels are vetoed.
The lepton \pt is required to be larger than 27~\GeV\ for the \olep channel;
for the \tlep channel, the leading~(subleading) lepton \pt must be larger than 28~(20)~\GeV,
and the invariant mass of the lepton pair must lie within $83 < m_{\ell\ell} < 99$~\GeV.
For the \zlep\ channel, the transverse momentum of the neutrino system must satisfy $\ptvv > 200$~\GeV;
and for the \olep\ channel, the events are required to have $\ptv > 80$~\GeV\ and contain no $b$-jets.
\begin{table}[h]
\begin{center}
\caption{Fiducial phase-space definitions used for the measurement of electroweak \VVjj\ production.
}
\label{tbl:vbs_fid_sel}
\begin{tabular}{lccc}
\toprule
\multicolumn{4}{c}{Object selection} \\
\midrule
Leptons      & \multicolumn{3}{c}{$\pt > 7$~\GeV, $|\eta| < 2.5$ } \\
Small-$R$ jets & \multicolumn{3}{c}{$\pt > 20$~\GeV\ if $|\eta| < 2.5$, and $\pt > 30$~\GeV\ if $2.5 < |\eta| < 4.5$} \\
Large-$R$ jets & \multicolumn{3}{c}{$\pt > 200$~\GeV, $|\eta| < 2.0$ } \\
\midrule
\multicolumn{4}{c}{Event selection} \\
\midrule
\multirow{4}{*}{Leptonic $V$ selection}
&  \zlep  & \multicolumn{2}{c}{ Zero leptons, $\ptvv > 200$~\GeV }
\\ \cline{2-4}
&  \olep  & \multicolumn{2}{c}{ One lepton with $\pt > 27$~\GeV, $\ptv > 80$~\GeV }
\\ \cline{2-4}
&  \multirow{2}{*}{\tlep}  & \multicolumn{2}{c}{ Two leptons, with leading~(subleading) lepton $\pt > 28~(20)$~\GeV} \\
&                            & \multicolumn{2}{c}{ $83<m_{\ell\ell}<99$~\GeV } \\
\midrule
\\ [-4mm]
\multirow{5}{*}{Hadronic $V$ selection}
& \multirow{2}{*}{Merged}
& \multicolumn{2}{c}{One large-$R$ jet, \minwzJ} \\
&         & \multicolumn{2}{c}{$64 < \mJ < 106$~\GeV } \\
\\[-2mm] \cline{2-4}
& \multirow{3}{*}{Resolved}
& \multicolumn{2}{c}{Two small-$R$ jets, \minwzjj} \\
& & \multicolumn{2}{c}{$p^{j_{1}}_\mathrm{T}>$40~\GeV, $p^{j_{2}}_\mathrm{T}>$20~\GeV } \\
& & \multicolumn{2}{c}{$64 < \mjj < 106$~\GeV } \\
\midrule
\\ [-4mm]
\multirow{2}{*}{Tagging-jets}
& \multicolumn{3}{c}{Two small-$R$ non-$b$ jets, $\etajo \cdot \etajt < 0$, highest \mjjtag} \\
& \multicolumn{3}{c}{$\mjjtag>400$~\GeV, $p^{\mathrm{tag},j_{1,2}}_\mathrm{T} > 30$~\GeV}  \\
\midrule
\\[-4mm]
\multirow{4}{*}{Number of $b$-jets}
&  \zlep  & \multicolumn{2}{c}{ -- }
\\ \cline{2-4}
&  \olep  & \multicolumn{2}{c}{ 0 }
\\ \cline{2-4}
&  \tlep  & \multicolumn{2}{c}{ -- }
\\
\bottomrule
 
\end{tabular}
\end{center}
\end{table}

% End of text imported from the .//./sections/FidXsec.tex input file
 
\section{Systematic uncertainties}
\label{sec:syst}
% The next lines are included from the .//./sections/Syst.tex input file
The sources of systematic uncertainty can be divided into three categories:
experimental uncertainties related to the detector or to the reconstruction algorithms,
uncertainties in the estimations of background contributions, and uncertainties in modeling the signal.
Unless stated otherwise, the uncertainties quoted below are the uncertainties in the quantities themselves,
not the impact on the analysis sensitivity.

The uncertainty in the integrated luminosity of the dataset is  2.1\%.
It is derived from the calibration of the luminosity scale using $x$-$y$ beam-separation scans,
following a methodology similar to that detailed in Ref.~\cite{DAPR-2013-01},
and using the LUCID-2 detector for the baseline luminosity measurements~\cite{LUCID2}.
This uncertainty is applied to the normalization of the signal and also to background contributions
whose normalizations are derived from MC simulations.
In addition to the luminosity uncertainty, a variation in the pileup reweighting of MC events is also included to cover the uncertainty in the ratio of the predicted to measured inelastic cross sections in Ref.~\cite{STDM-2015-05}.
 
The efficiencies of the lepton triggers for events with selected leptons are high, nearly 100\% in the electron channel
and approximately 96\% in the muon channel.  The corresponding uncertainties are negligible.
For the selection used in the \zlep and \olep channels, the efficiency of the $\met$ trigger is also close to 100\%
with negligible associated uncertainty.
The modeling of the electron and muon reconstruction, identification and isolation efficiencies
is studied with a tag-and-probe method using $Z\to\ell\ell$ events in data and simulation at \cms\
\cite{PERF-2015-10,PERF-2017-01}.
Small corrections are applied to the simulation to better model the performance seen in data.
These corrections have associated uncertainties of the order of 1\%.
Uncertainties in the lepton energy (or momentum) scale and resolution~\cite{PERF-2015-10,PERF-2017-03} are also taken into account.

Uncertainties in the jet energy scale and resolution for small-radius jets
are estimated using MC simulation and {\em in situ} techniques~\cite{PERF-2016-04}.
For central jets ($|\eta|<2.0$), the total uncertainty in the jet energy scale ranges from about 6\%
for jets with $\pt = 25~\gev$ to about 2\% for $\pt = 1~\tev$.
There is also an uncertainty in the jet energy resolution~\cite{PERF-2016-04},
which ranges from 10\% to 20\% for jets with a $\pt$ of $20~\gev$ to less than 5\% for jets with $\pt>200~\gev$.
Uncertainties in the lepton and jet energy scales and resolutions are propagated into
the uncertainty in $\met$.
Uncertainties in the energy scale and resolution of the track soft term
are also propagated into the uncertainty in $\met$~\cite{PERF-2016-07}.
For the $b$-tagging efficiency of small-$R$ jets, correction factors are applied to the simulated event samples
in order to compensate for differences between data and simulation.
The corrections and uncertainties in the efficiency for tagging $b$-jets and in the rejection factor
for light jets are determined from $\ttbar$ samples~\cite{ATL-PHYS-PUB-2015-039,ATL-PHYS-PUB-2015-022}.
 
The uncertainties in the scale of the large-$R$ jet $\pT$, mass and \DTwoBetaOne\
are of the order of 2--5\%.
They are estimated using comparisons of data and simulation in Ref.~\cite{JETM-2018-03}.
An absolute uncertainty of 2\% is assigned to the large-$R$ jet energy resolution, and relative uncertainties of 20\% and 15\% are assigned to the resolution of the large-$R$ jet mass and \DTwoBetaOne, respectively.

The overall normalization of the main backgrounds (\Wjets, \Zjets and \ttbar)
is determined from the corresponding data control regions and is left unconstrained and floating in the global likelihood fit.
For \Wjets (\Zjets) events in the \zlep channel, additional normalization uncertainties are considered
to account for the acceptance difference between the \zlep channel analysis and the \olep (\tlep) channel analysis,
given that there are no corresponding pure control regions of \zlep\ events and
the normalization is determined mainly from control regions with \olep (\tlep) events.
This additional normalization uncertainty for \Wjets (\Zjets) events is estimated using
the ratio of the event yield in each signal region of the \zlep channel to that in the \olep (\tlep) channel,
and by comparing this ratio obtained from the nominal MC samples generated by \textsc{Sherpa} with the ratio from
alternative samples generated by \MGMCatNLO.
The normalization uncertainty is 8\%~(14\%) for \Wjets events in the merged (resolved) signal region,
and 22\%~(42\%) for \Zjets events in the merged (resolved) signal region.
These uncertainties are applied to the \Wjets and \Zjets events in the \zlep channel only.
The normalization uncertainties in the diboson background cross sections are studied with \textsc{Sherpa}.
The uncertainty due to missing higher-order QCD contributions~(QCD scale uncertainty)
is estimated by varying the renormalization~(\muR) and factorization~(\muF) scales independently by a factor
ranging from one-half to two with the constraint $0.5 \leq \muF/\muR \leq 2$.
The PDF uncertainty corresponds to the 68\% confidence-level variations of the nominal PDF set \textsc{NNPDF30NNLO},
as well as its difference from the alternative PDF sets CT10NNLO~\cite{CT10NNLO} and MMHT2014NNLO~\cite{MMHT2014}.
The overall normalization uncertainty for the diboson background is estimated to be about 30\%.
For single-top-quark events, a 20\% normalization uncertainty is assigned~\cite{TOPQ-2015-16}.

The uncertainty in the modeling of the final discriminants, the BDT output and \mjjtag, for background processes estimated using
MC simulation is assessed by comparing the nominal MC samples with alternative samples.
The uncertainties are of the order of 5--30\%.
The \mjjtag reweighting as described in Section~\ref{sec:bkgCR}
is also included as a shape systematic uncertainty for \Zjets and \Wjets events
by taking the difference of their respective final discriminants before and after applying the reweighting.
An uncertainty in the shape of the  BDT or \mjjtag
distribution for the \ttbar  background is derived by comparing the \textsc{Powheg-Box} sample with the
distribution obtained using \MGMCatNLO~2.2.2.
Additional systematic uncertainties are estimated by comparing the
nominal sample showered with \textsc{Pythia}~6.428 using the P2012 tune to
one showered with \Herwigpp~2.7.1~\cite{Herwig} and using the UEEE5 underlying-event tune~\cite{Gieseke:2012ft}.
Samples of \ttbar events with the factorization and renormalization
scales doubled or halved are compared with the nominal samples, and the observed differences are taken as an additional uncertainty.
These modeling uncertainties for the \ttbar background are 5--30\%.
The shape uncertainty for diboson processes is obtained by comparing MC samples generated by
\textsc{Sherpa} and \textsc{Powheg-Box}, and it is found to be of the order of 2--30\%.
The shape uncertainty for single-top-quark events is ignored due to their relatively small contribution to the total
background.

The following discussion describes the uncertainties in the predictions of EW \VVjj\ signal processes.
The uncertainties in the signal-strength measurement, discussed in Section~\ref{sec:res_mu}, include contributions from
both the normalization and shape; for the fiducial cross section measurement, discussed  in Section~\ref{sec:res_fidxs}, only the shape
uncertainties are taken into account for the measured fiducial cross sections, and the normalization uncertainties are
included for the SM predicted fiducial cross sections.

Theoretical uncertainties for EW \VVjj\ signal processes include the PDF choice, the missing higher-order
corrections, and the parton-shower modeling.
The signal modeling uncertainty due to PDF uncertainties is estimated by taking the uncertainty
from the PDF error sets of NNPDF23LO and adding it in quadrature to the acceptance difference obtained using
alternative PDF sets: \textsc{CT10} and MMHT2014LO.
The PDF uncertainties are estimated to be 3--5\%.
The parton-shower uncertainty, estimated by varying relevant parameters in the A14-NNPDF tune~\cite{ATL-PHYS-PUB-2014-021},
ranges from 1\% to 5\%.
The effect of the QCD scale uncertainty, of the order of 1--3\%, is estimated by varying the factorization and renormalization
scales independently by a factor of two with the constraint $0.5 \leq \muF/\muR \leq 2$.

The interference between EW- and QCD-induced \VVjj\ processes
is not included in the MC simulation, since the EW- and QCD-induced \VVjj\ samples were generated separately. The interference effect is
considered as an uncertainty affecting both the normalization and the shape of the EW \VVjj\ kinematic distributions.
The effect is determined using the \MGMCatNLO~2.4.3 MC generator at the `truth' level as a function of \mjjtag.
A reweighting is then applied to the simulated EW \VVjj\ samples,
resulting in shape uncertainties of 5\% to 10\% at low and high values of the BDT score, respectively,
and a similar size for the normalization uncertainties.
% End of text imported from the .//./sections/Syst.tex input file
\section{Statistical analysis}
\label{sec:fit}
% The next lines are included from the .//./sections/Fit.tex input file
The statistical analysis relies on the profile likelihood test statistic~\cite{HIGG-2012-17} implemented with the
RooFit~\cite{Verkerke:2003ir} and RooStats~\cite{Moneta:2010pm} packages.
A binned likelihood function $\mathcal{L}(\mu,\theta)$ is constructed as a product of Poisson probabilities over all of the bins
of the fit templates considered in the analysis. This function depends on the signal-strength parameter $\mu$,
a multiplicative factor applied to the theoretical signal production cross section, and $\theta$,
a set of nuisance parameters that encodes the effects of systematic uncertainties in the signal and expected backgrounds.
The binning is chosen so that the expected numbers of events ensure that
the statistical uncertainty is less than 5\% in most bins, while finer binning is also allowed in signal-enriched regions.
The nuisance parameters are either free to float, or constrained using Gaussian or log-normal terms defined by external studies.
The likelihood function for the combination of the three channels is the product of the Poisson likelihoods of the
individual channels. However, only one constraint term per common nuisance parameter is included in the product.

A simultaneous maximum-likelihood fit is performed to the observed distributions
of the final discriminants, BDT outputs, in the nine SRs to extract the signal rate information.
The three ZCRs, WCRs and TopCRs as well as the three VjjCRs are included in the fit's likelihood calculation;
the \mjjtag\ distributions are used for ZCRs, WCRs and VjjCRs, while for the TopCRs
only one bin for each of the three \Vhad\ decay channels is used.
The purpose of using \mjjtag\ distributions for CRs is to constrain the \mjjtag\ reweighting
systematic uncertainties.
The different regions and the corresponding discriminants entering the likelihood fit are summarized in Table~\ref{tab:fitregions_all}.
Signal and background contributions, including their shapes in the signal and control regions, are taken from MC simulations.
For each source of systematic uncertainty, the correlations across bins of BDT distributions are taken into account
and are fully correlated.
The correlations between different regions, as well as those between signal and background, are also included.
Moreover, normalization scale factors~(SFs) are applied to the MC estimates of the $Z$+jets, $W$+jets and
top quark contributions.
These SFs are free parameters in the fit and are therefore constrained by the data in both the signal
and control regions.
The diboson contribution is constrained to the theoretical estimate within the corresponding uncertainties.
 
\begin{table}[htb!]
\centering
\caption{\label{tab:fitregions_all}
The distributions used in the global likelihood fit for the signal regions and control regions for all the categories
in each channel.
``One bin'' implies that a single bin without any shape information is used in the corresponding fit region.
}
\begin{tabular}{llccc}
\toprule
\multicolumn{2}{c}{\multirow{2}{*}{Regions}} & \multicolumn{3}{c}{Discriminants} \\
\cmidrule{3-5}
& & Merged high-purity & Merged low-purity & Resolved \\
\midrule
\multirow{2}{*}{\zlep}
& SR      & BDT  & BDT & BDT \\
& VjjCR   & \mjjtag  & \mjjtag & \mjjtag \\
\midrule
\multirow{3}{*}{\olep}
& SR     & BDT & BDT & BDT \\
& WCR    & \mjjtag & \mjjtag & \mjjtag \\
& TopCR  & One bin & One bin & One bin \\
\midrule
\multirow{2}{*}{\tlep}
& SR     & BDT & BDT & BDT \\
& ZCR    & \mjjtag & \mjjtag & \mjjtag \\
\bottomrule
\end{tabular}
\end{table}

In general, one SF is introduced for each background component, common to both the SRs and CRs.
One common \Zjets SF is used for both the \zlep and \tlep channels,
and one common \Wjets SF is used for both the \zlep and \olep channels.
Similarly, one common \ttbar SF is used for both the \zlep and \olep channels.
However, independent SFs are used for the resolved and merged categories,
to take into account different MC modelings in the different phase spaces of the same background component.

The test statistic $q_{\mu}$ is defined as the profile likelihood ratio~\cite{ATL-PHYS-PUB-2011-011},
$q_{\mu}=-2\ln\Lambda_{\mu}$ with $\Lambda_{\mu}=\mathcal{L}(\mu,\hat{\hat{\theta}}_{\mu})/ \mathcal{L}(\hat{\mu}, \hat{\theta})$,
where $\hat{\mu}$ and $\hat{\theta}$ are the values of the parameters that maximize the likelihood function
(with the constraint 0$\leq  \hat{\mu} \leq \mu$), and $\hat{\hat{\theta}}_{\mu}$ are the values of the
nuisance parameters that maximize the likelihood function for a given value of $\mu$.
The best-fit signal strength $\hat{\mu}$ value~(\muVBSobs) is obtained by maximizing the likelihood function with respect to all parameters.
To determine whether the observed data is compatible with the background-only hypothesis,
a test statistic $q_{0}=-2\ln\Lambda_{0}$ is used.

% End of text imported from the .//./sections/Fit.tex input file
 
\section{Results}
\label{sec:results}
% The next lines are included from the .//./sections/Results.tex input file
\subsection{Results for the EW \VVjj\ production processes}
\label{sec:res_mu}
\begin{figure}[hp]
\centering
\begin{tabular}{cccc}
\includegraphics[width=0.34\linewidth]{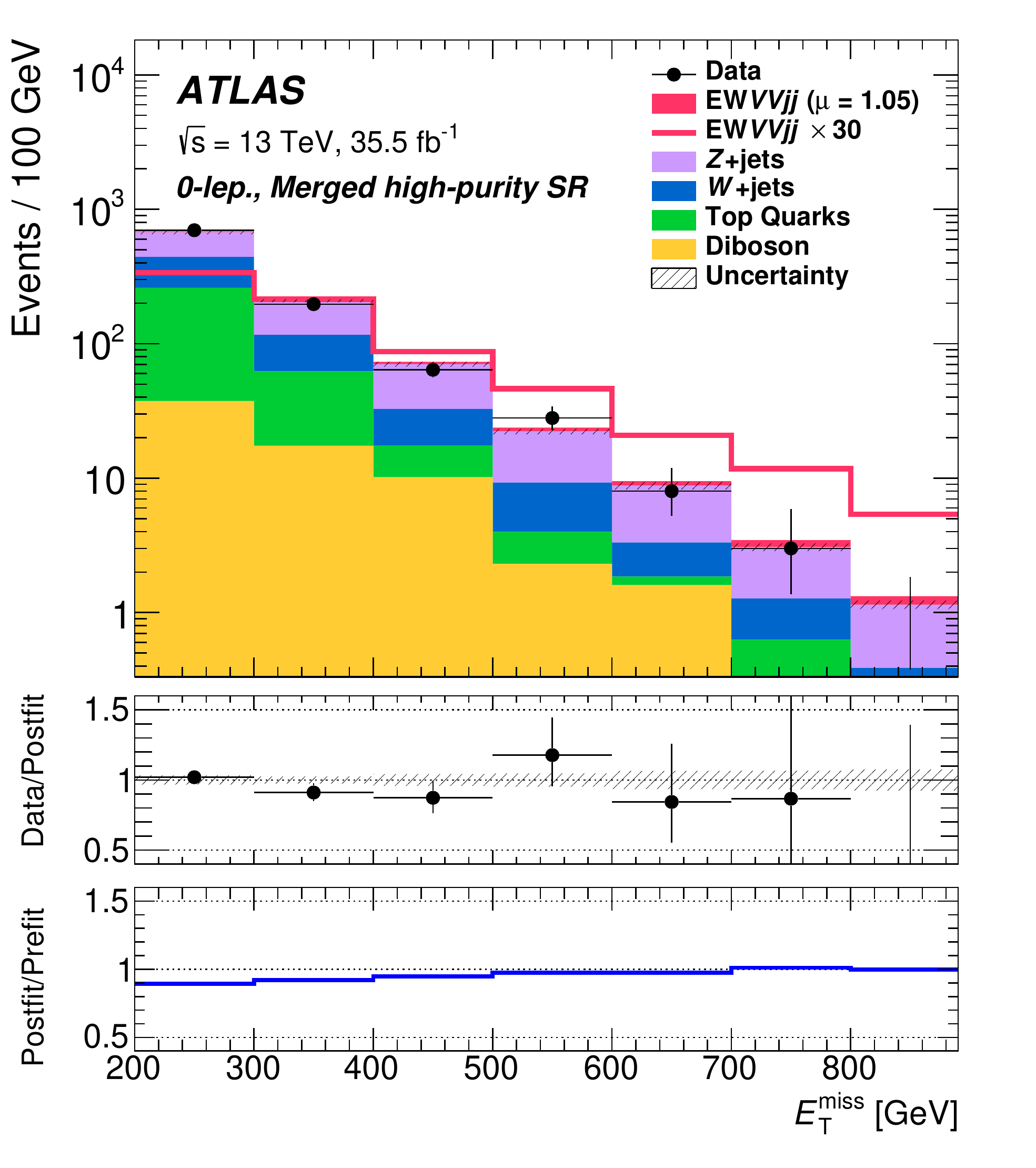}
\includegraphics[width=0.34\linewidth]{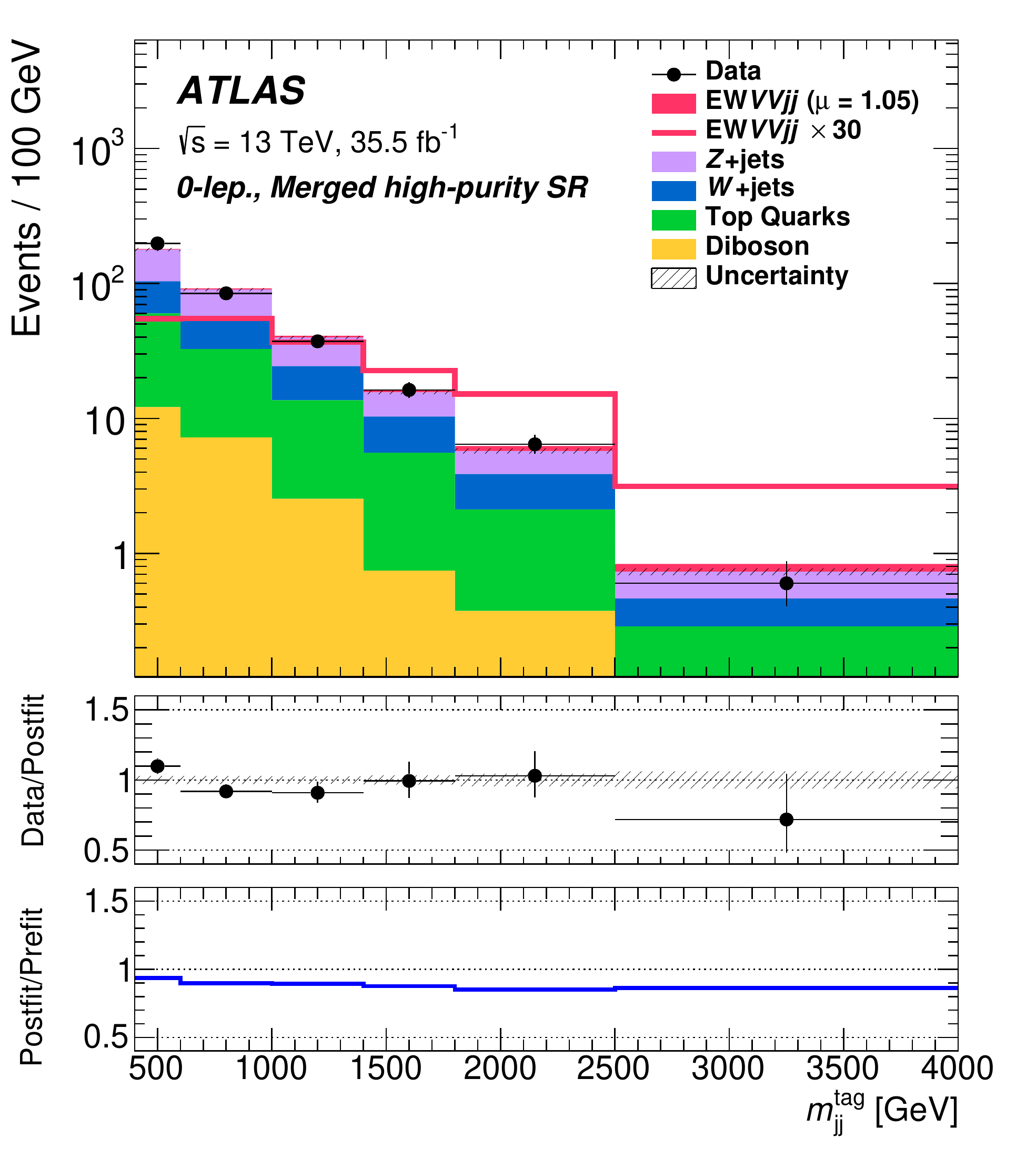}\\
\includegraphics[width=0.34\linewidth]{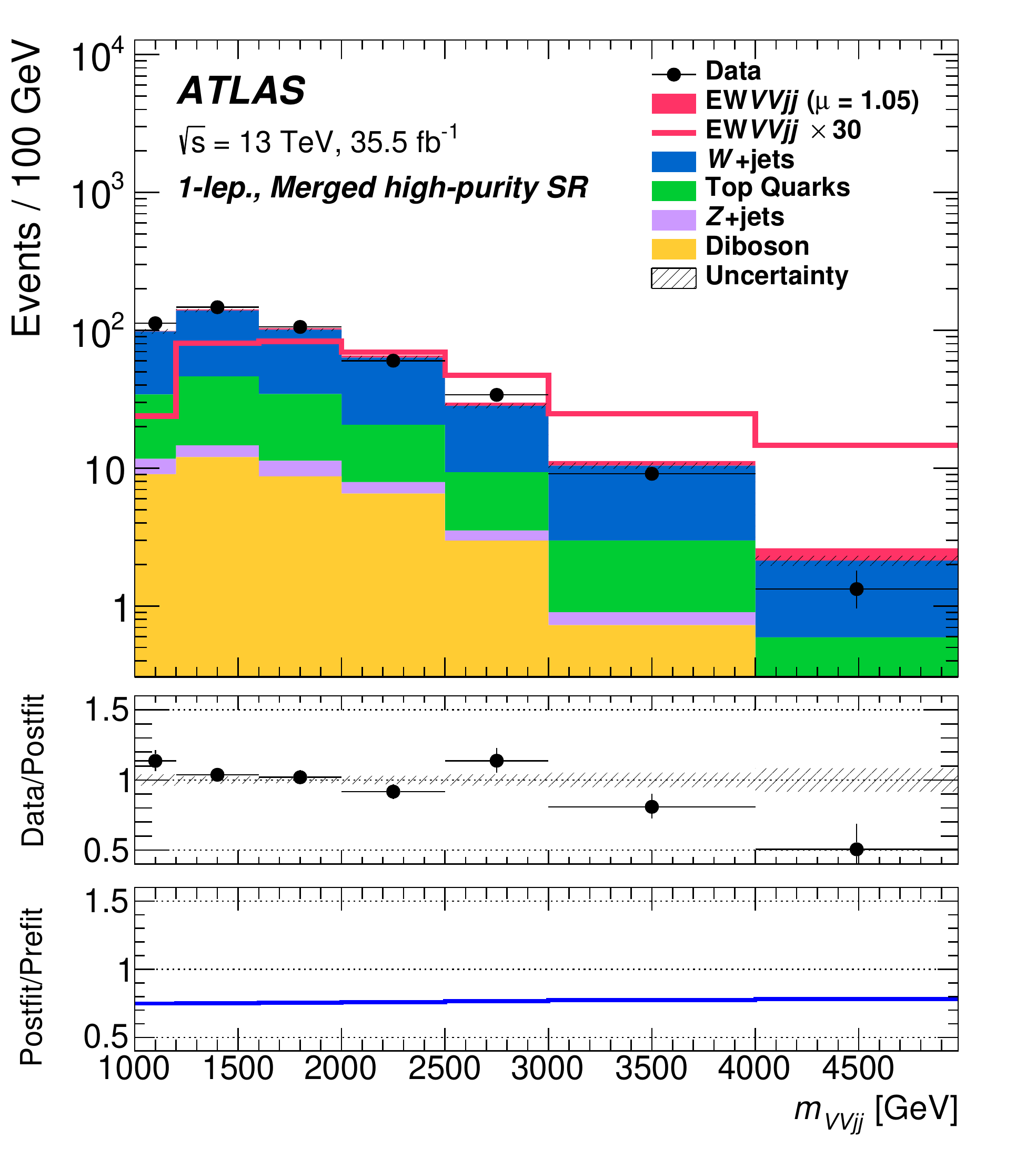}
\includegraphics[width=0.34\linewidth]{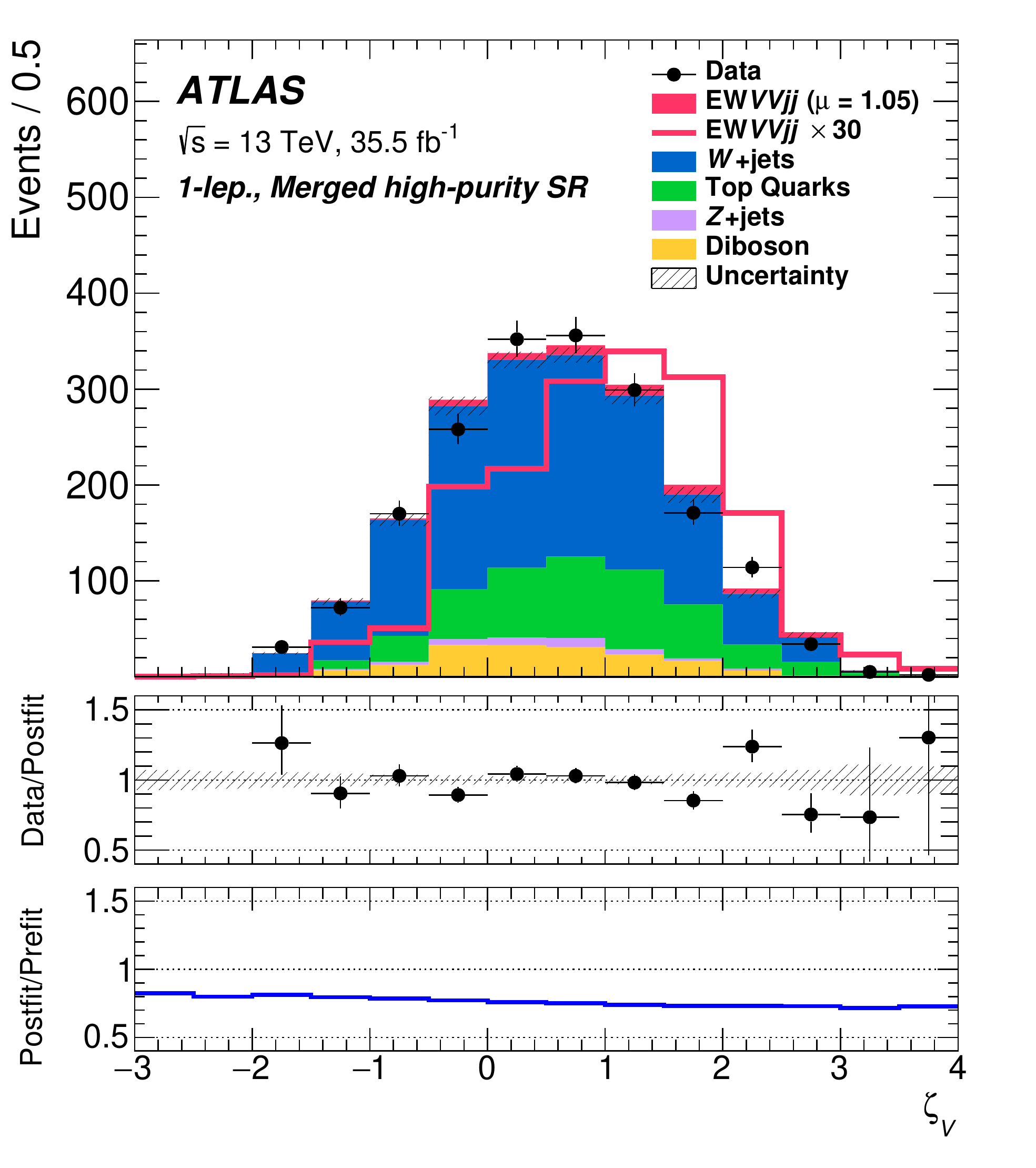}\\
\includegraphics[width=0.34\linewidth]{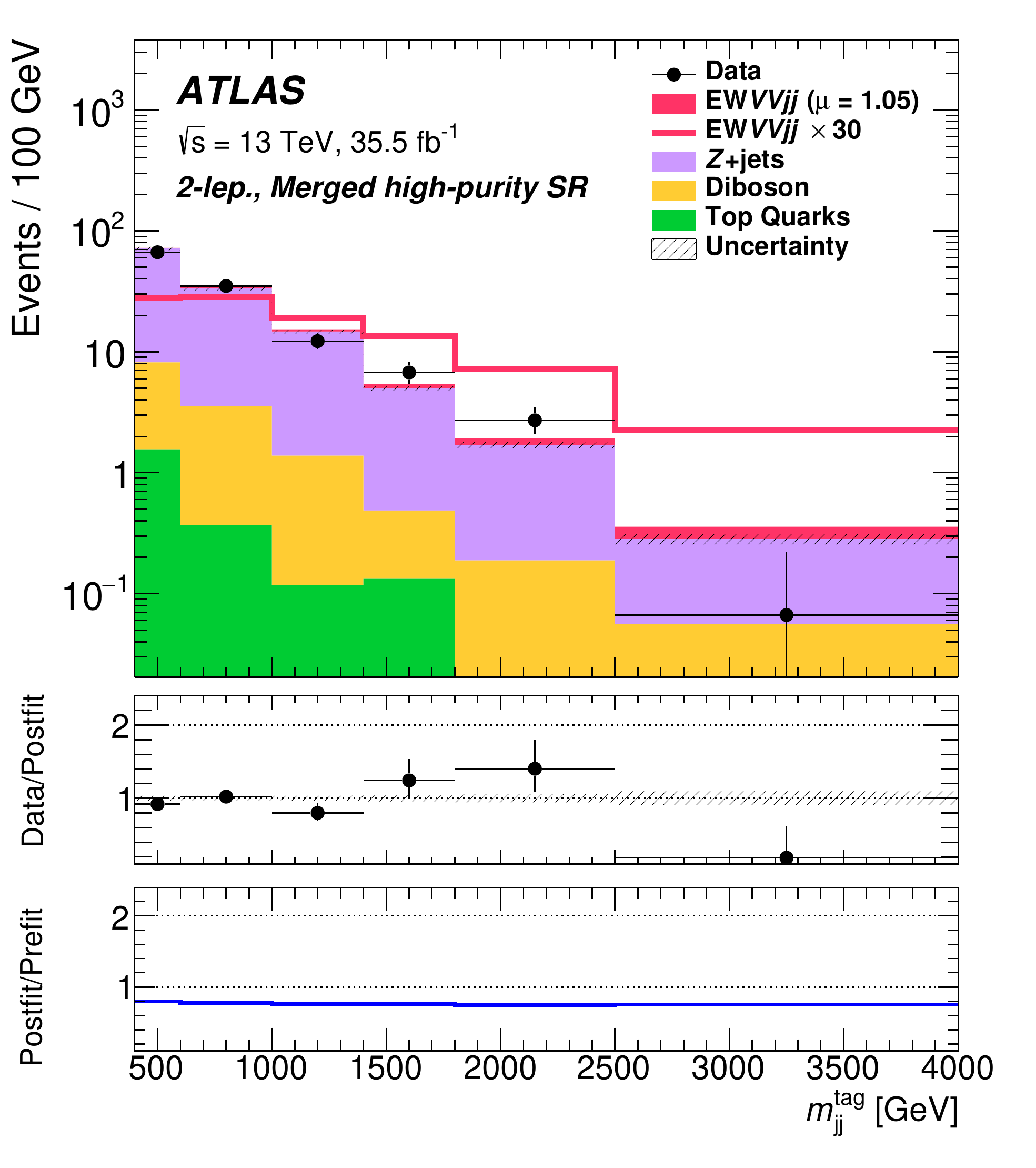}
\includegraphics[width=0.34\linewidth]{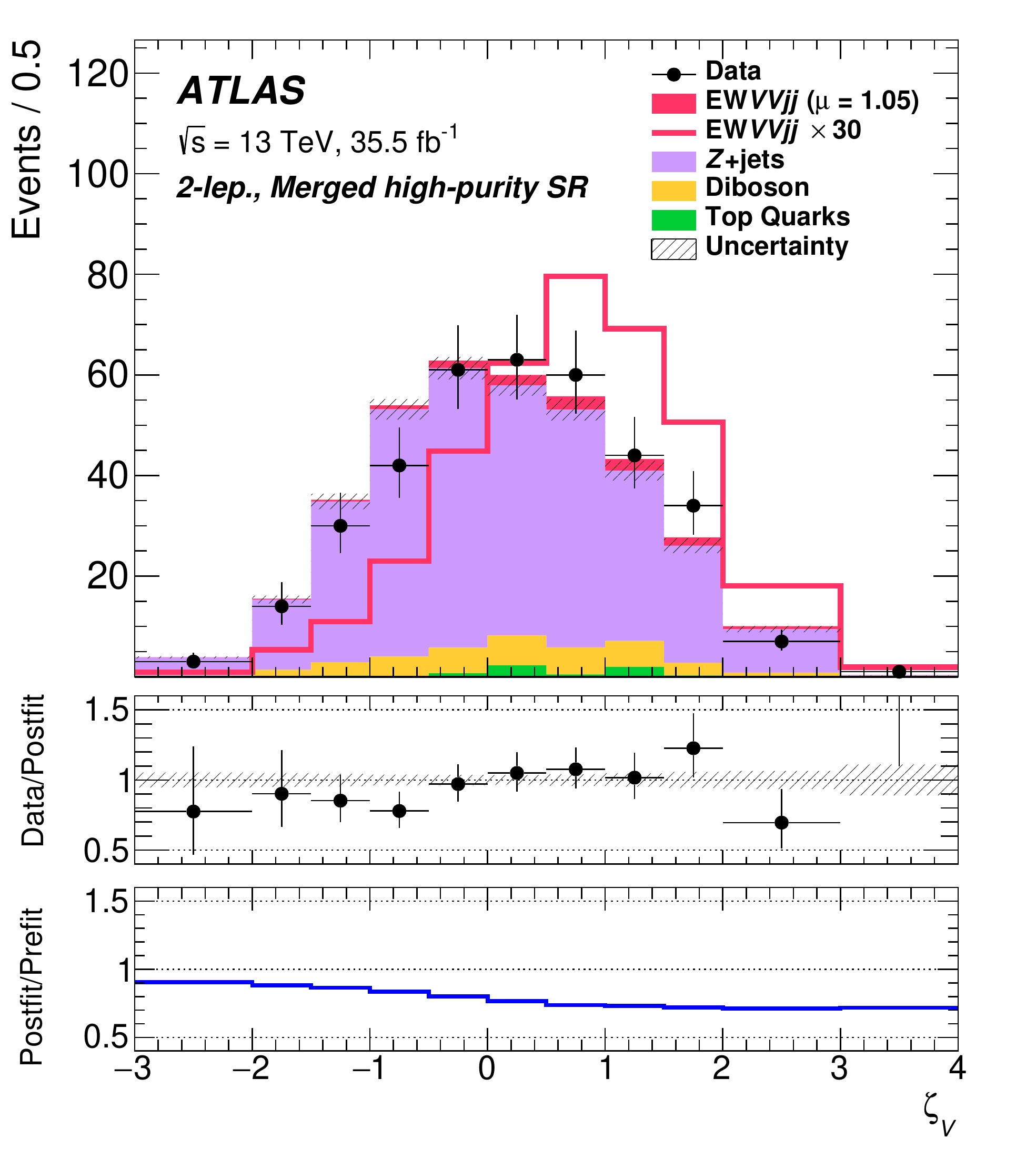}\\
\end{tabular}
\caption{The distributions for
\met~(top left), \mjjtag~(top right),
\mVVjj~(middle left), \zetaV~(middle right),
\mjjtag~(bottom left), and \zetaV~(bottom right)
in the 0-lepton (top), 1-lepton (middle) and 2-lepton (bottom) channels for the high-purity merged signal region.
The background contributions after the global likelihood fit are shown as filled histograms.
The signal is shown as a filled histogram on top of
the fitted backgrounds normalized to the signal yield extracted from data ($\mu=\muVBSobsNom$),
and unstacked as an unfilled histogram, scaled by the factor indicated in the legend.
The size of the combined statistical and systematic uncertainty for the
sum of the fitted signal and background is indicated by the hatched band.
The middle pane shows the ratios of the observed data to the post-fit signal and background predictions.
The bottom pane shows the ratios of the post-fit and pre-fit background predictions.
}
\label{fig:BDT_inputs_merged}
\end{figure}

\begin{figure}[hp]
\centering
\begin{tabular}{cccc}
\includegraphics[width=0.34\linewidth]{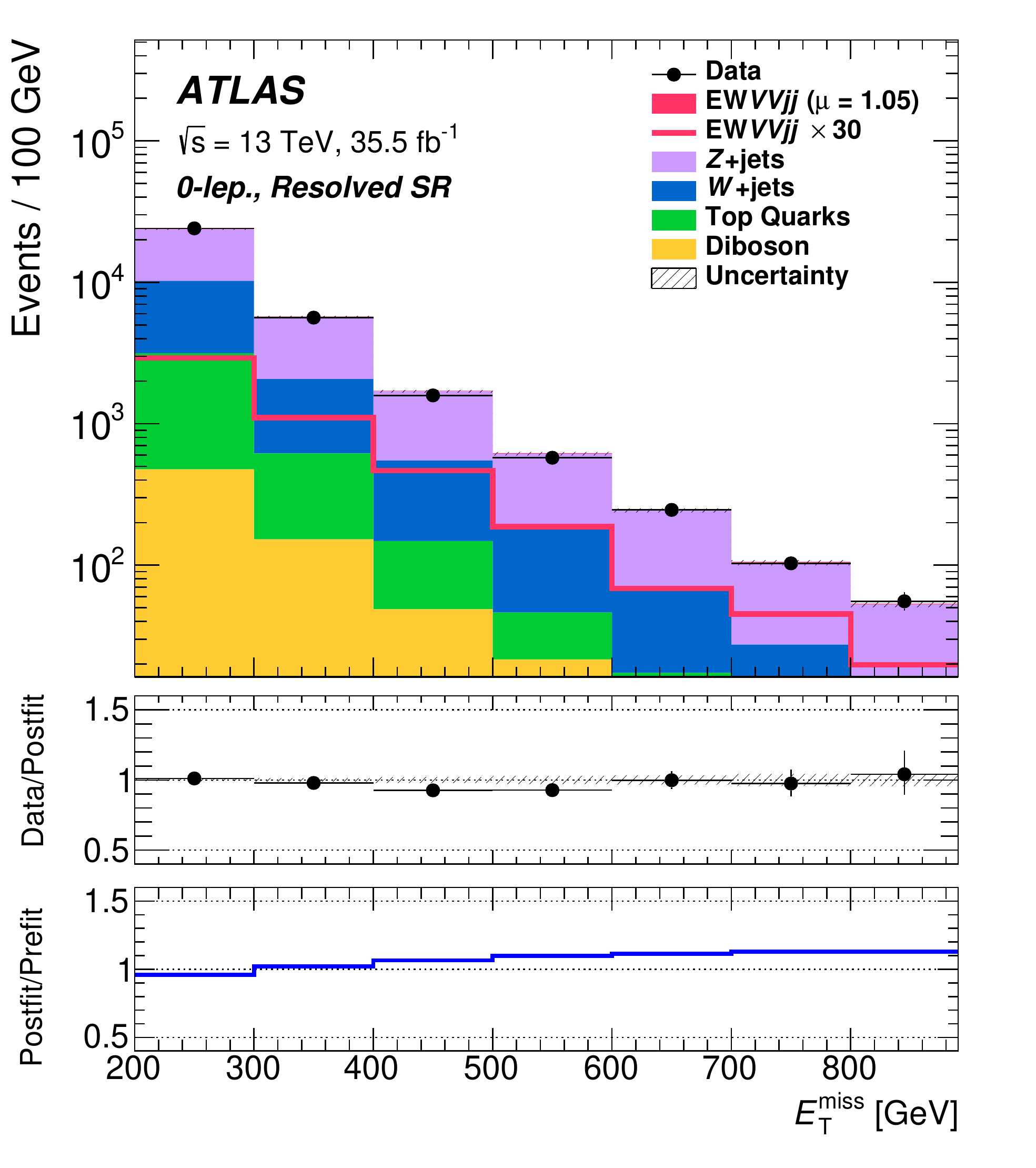}
\includegraphics[width=0.34\linewidth]{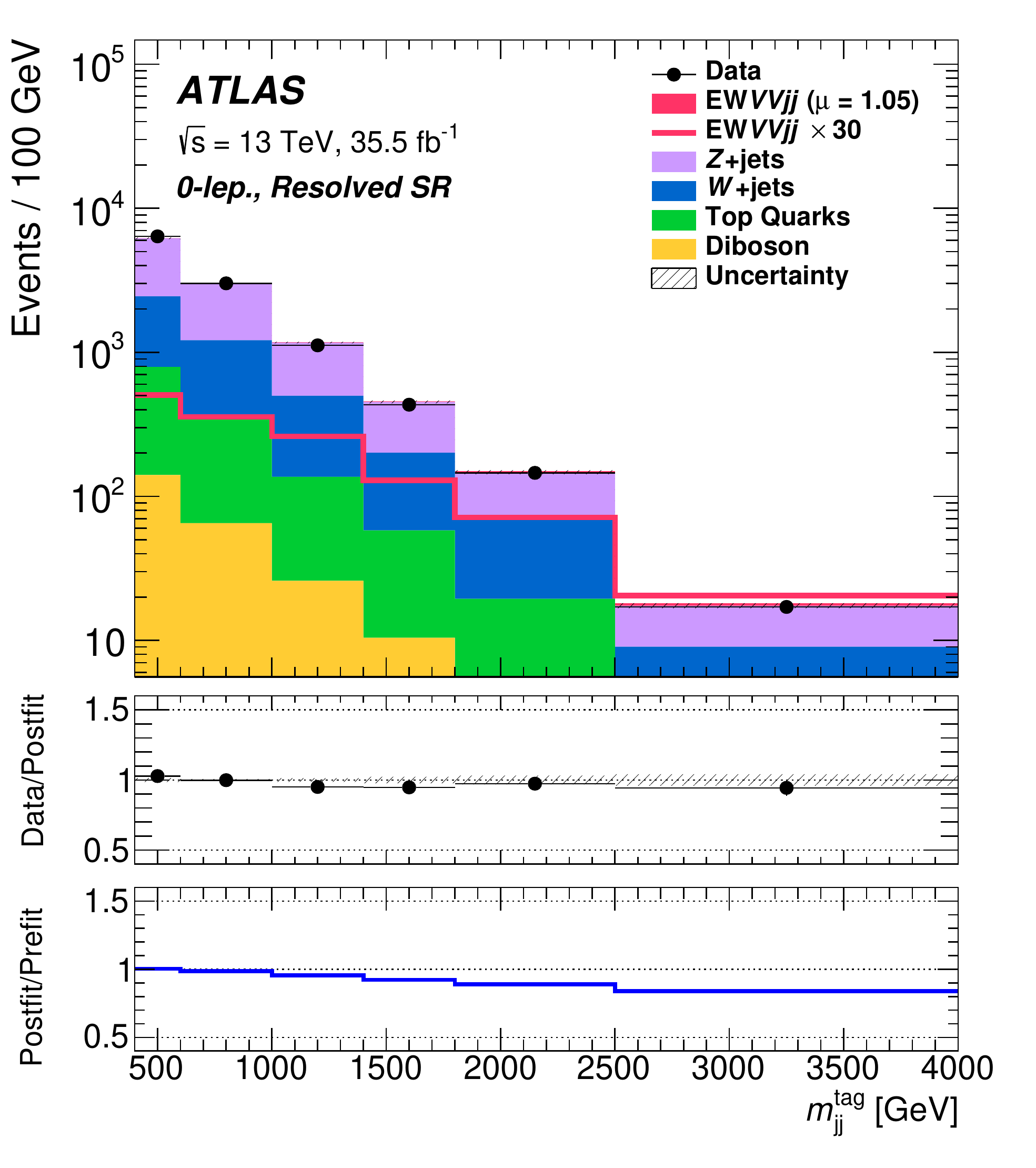}\\
\includegraphics[width=0.34\linewidth]{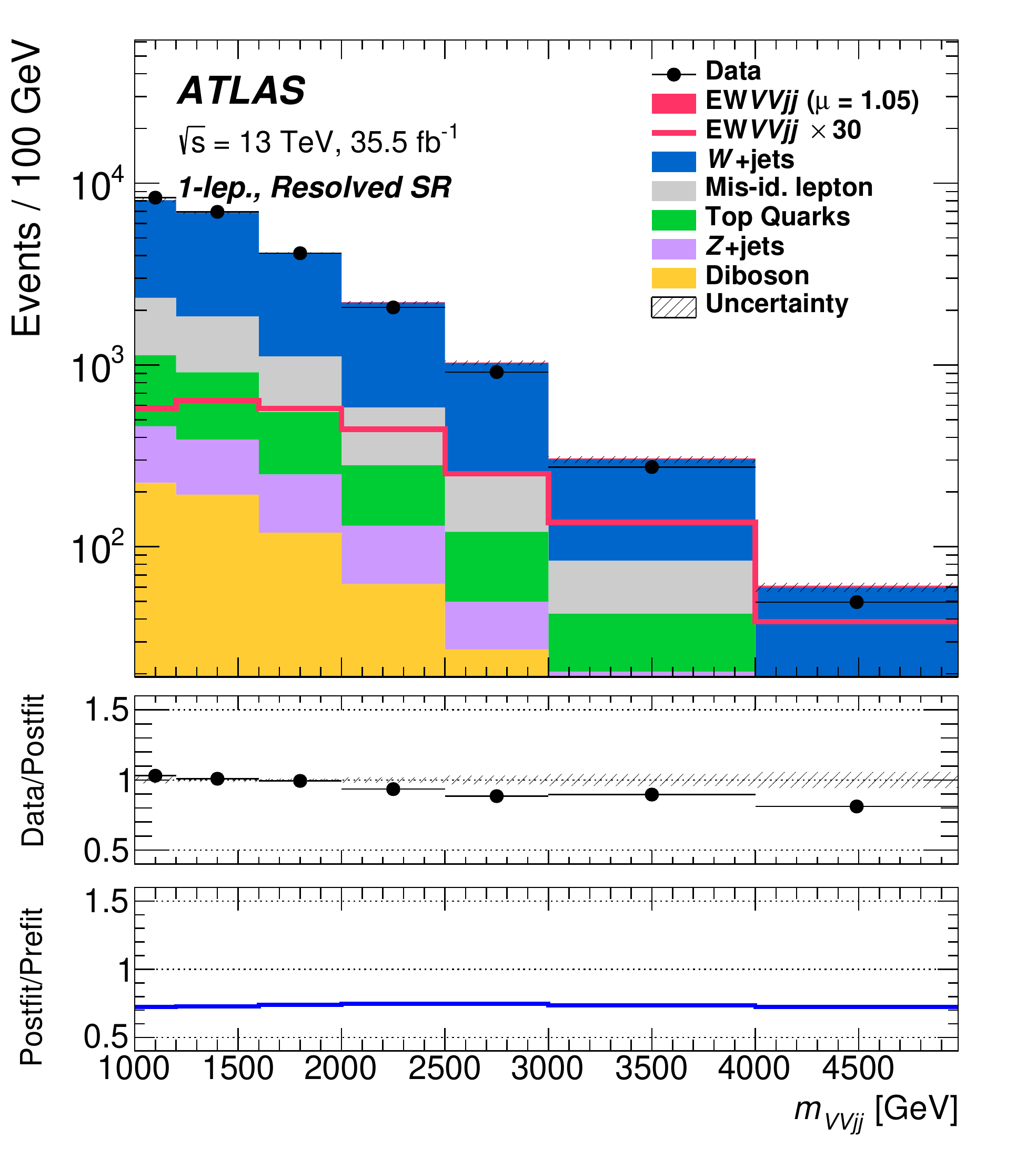}
\includegraphics[width=0.34\linewidth]{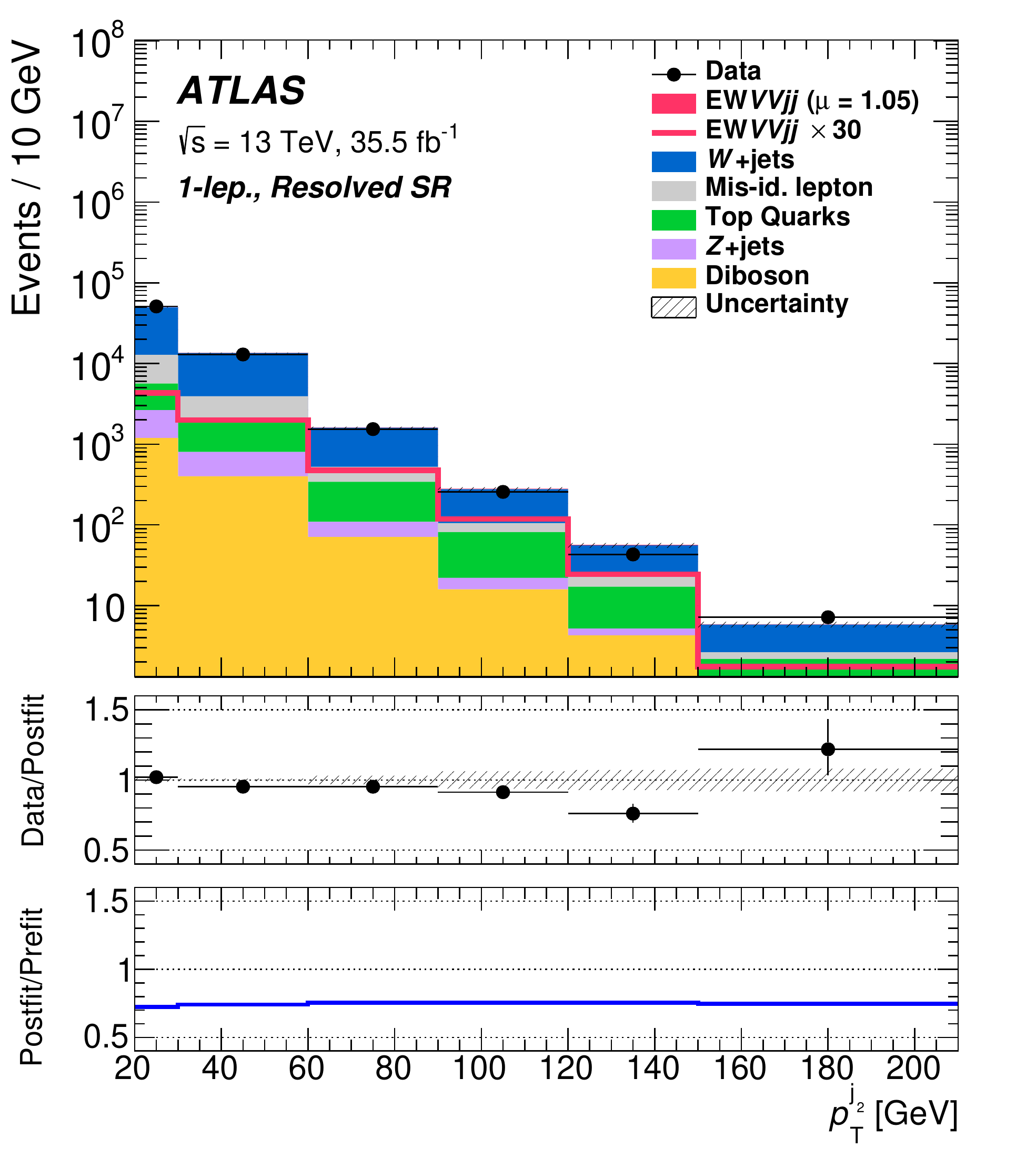}\\
\includegraphics[width=0.34\linewidth]{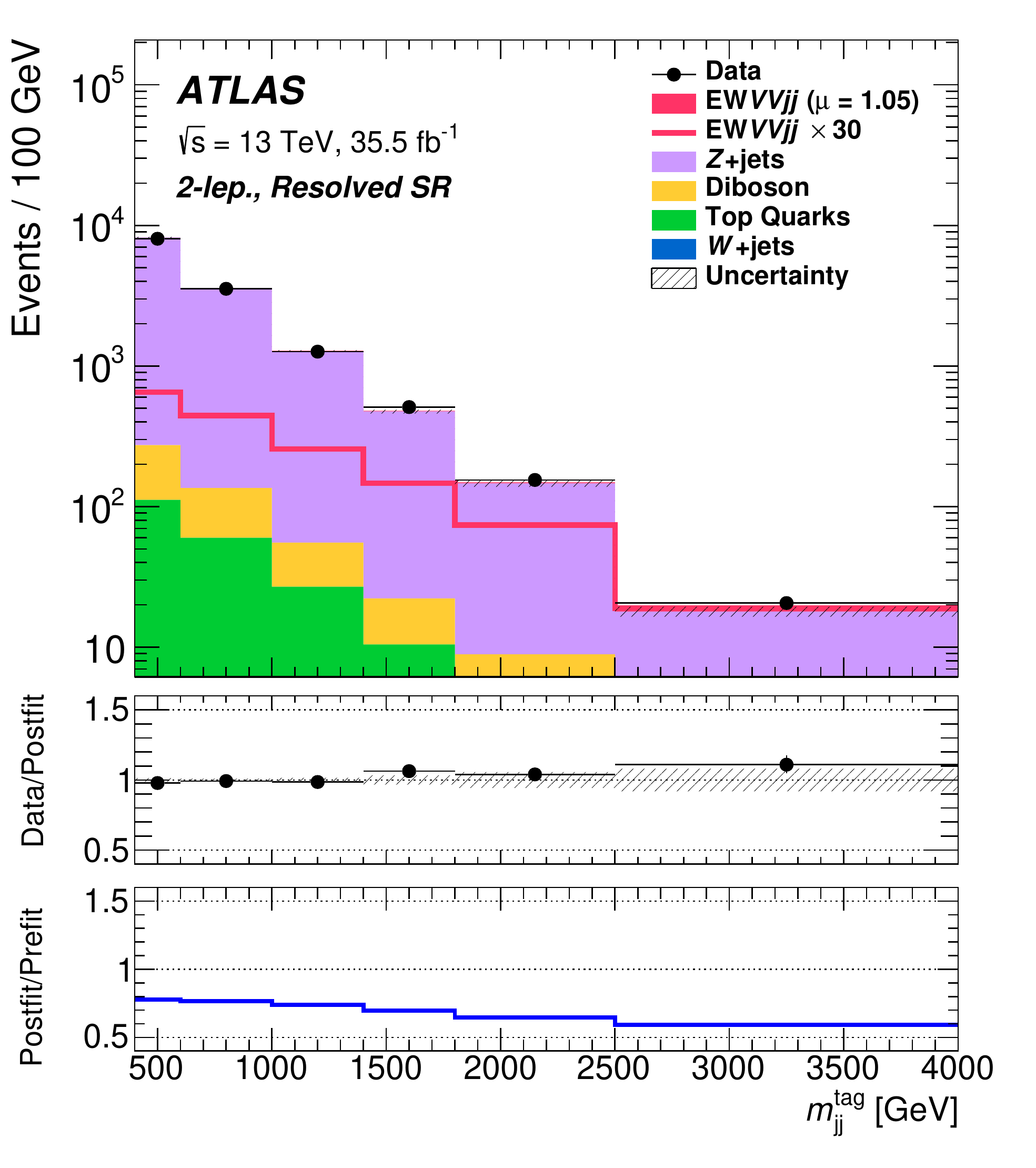}
\includegraphics[width=0.34\linewidth]{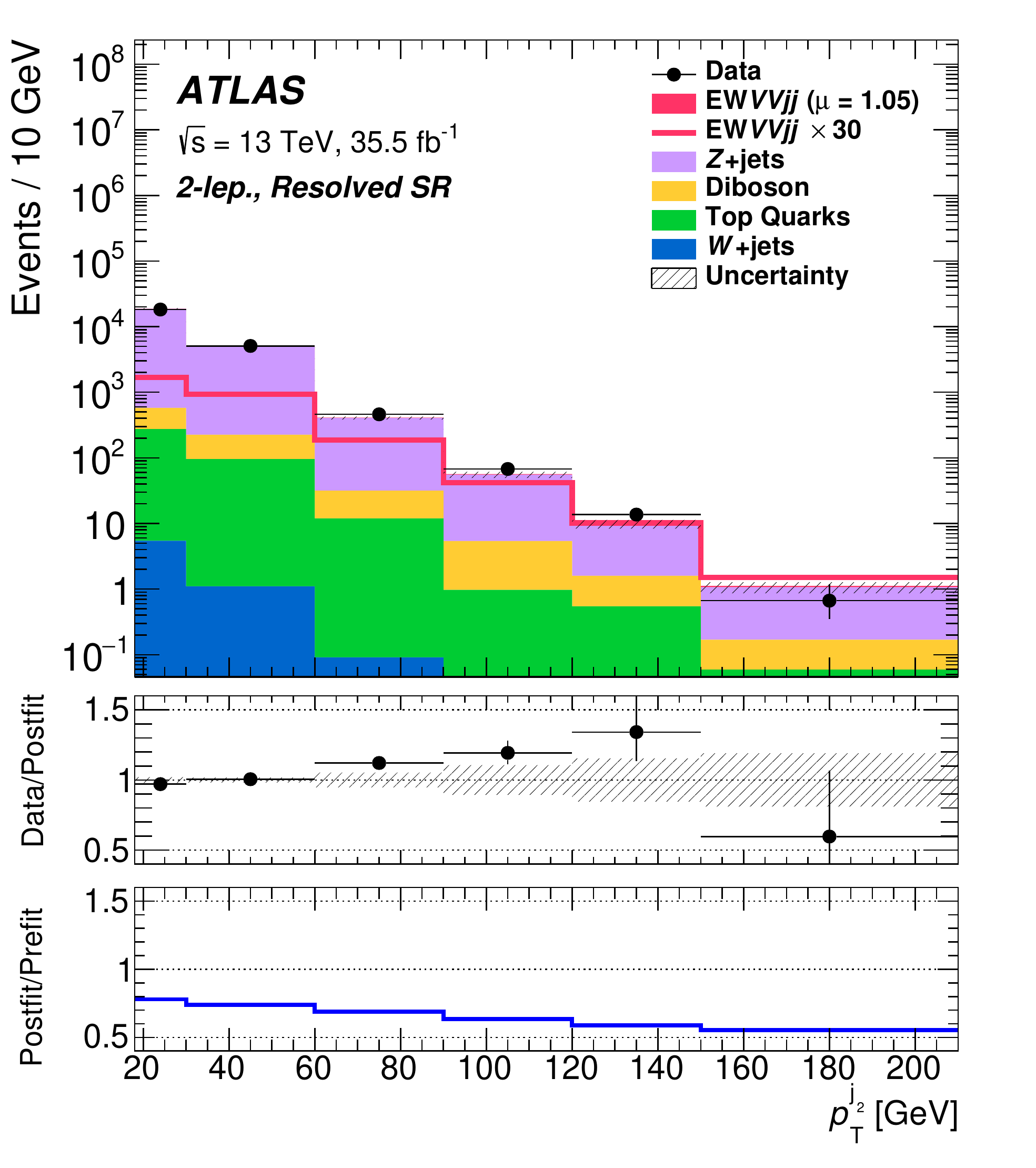}\\
\end{tabular}
\caption{The distributions for
\met~(top left), \mjjtag~(top right),
\mVVjj~(middle left), \ptjt~(middle right),
\mjjtag~(bottom left), and \ptjt~(bottom right)
in the 0-lepton (top), 1-lepton (middle) and 2-lepton (bottom) channels for the resolved signal region.
The background contributions after the global likelihood fit are shown as filled histograms.
The signal is shown as a filled histogram on top of
the fitted backgrounds normalized to the signal yield extracted from data ($\mu=\muVBSobsNom$),
and unstacked as an unfilled histogram, scaled by the factor indicated in the legend.
The size of the combined statistical and systematic uncertainty for the
sum of the fitted signal and background is indicated by the hatched band.
The middle pane shows the ratios of the observed data to the post-fit signal and background predictions.
The bottom pane shows the ratios of the post-fit and pre-fit background predictions.
}
\label{fig:BDT_inputs_resolved}
\end{figure}

Figures~\ref{fig:BDT_inputs_merged}  and \ref{fig:BDT_inputs_resolved}
show a selection of representative post-fit distributions of input variables
that are most discriminating for each of the lepton channels, for the merged and resolved categories, respectively.
Background and EW \VVjj\ signal contributions shown are obtained from the signal-plus-background fits described previously.

\begin{figure}[!h!btp]
\centering
\subfigure[]{\includegraphics[width=0.45\textwidth]{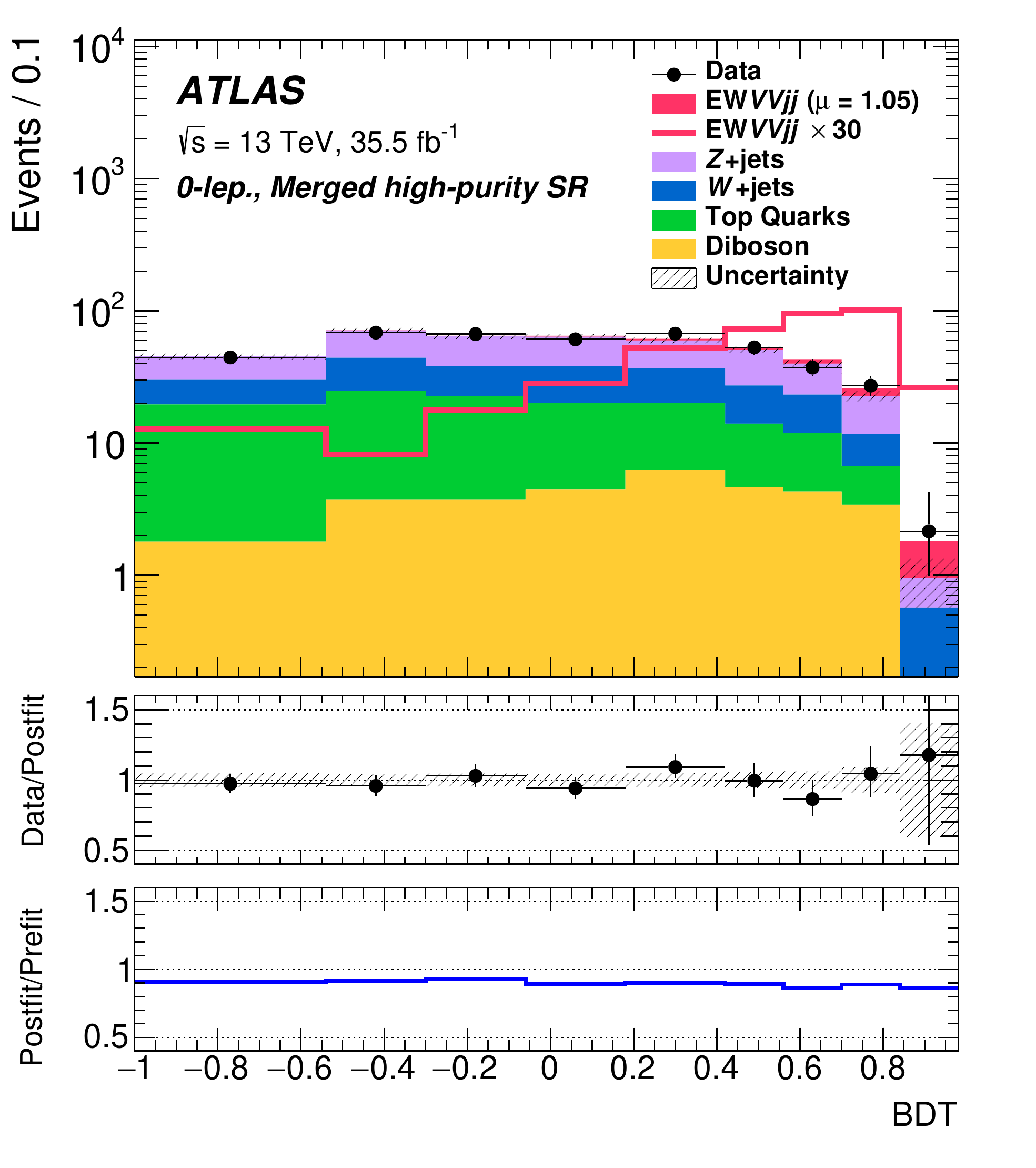}}
\subfigure[]{\includegraphics[width=0.45\textwidth]{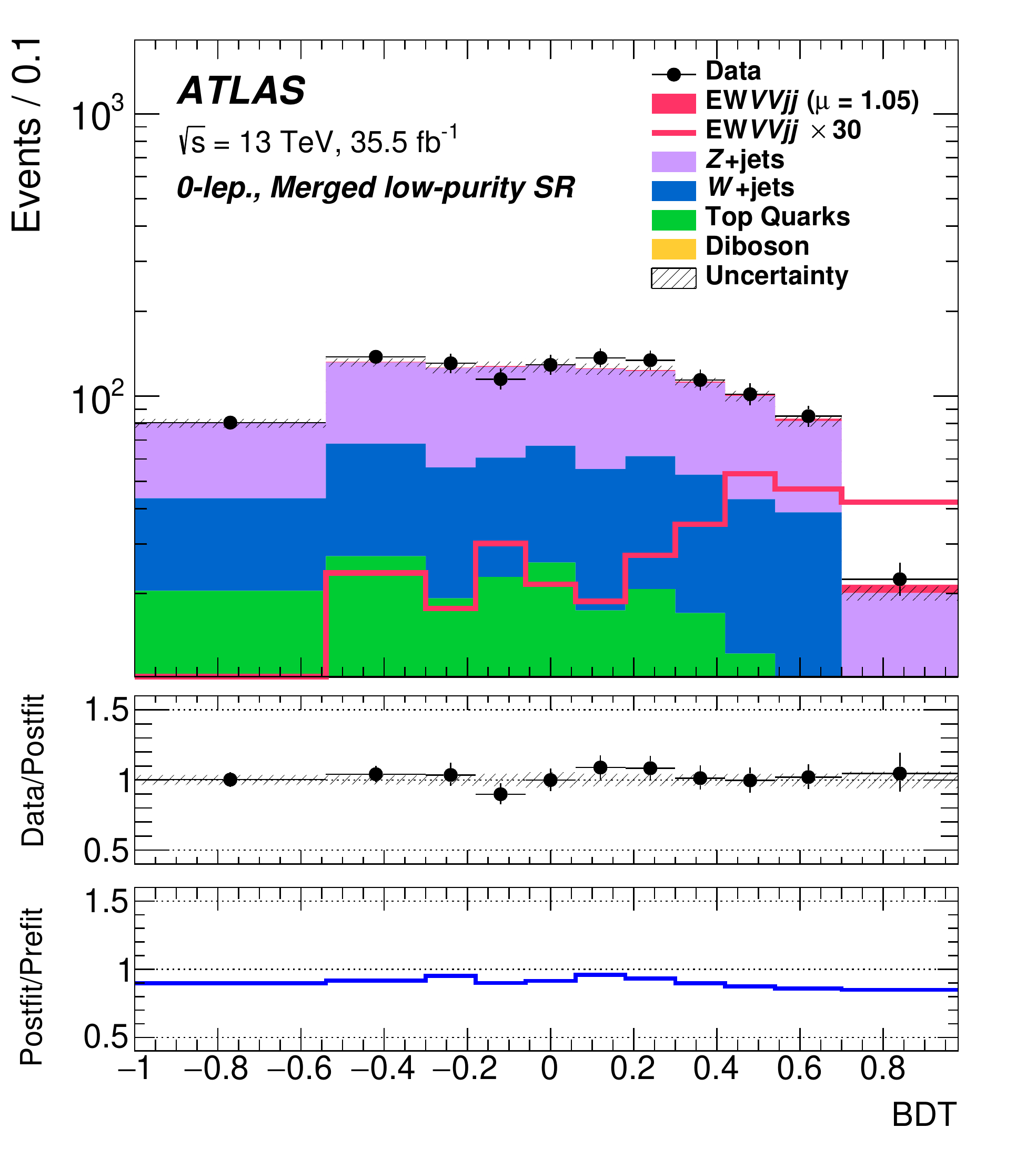}} \\
\subfigure[]{\includegraphics[width=0.45\textwidth]{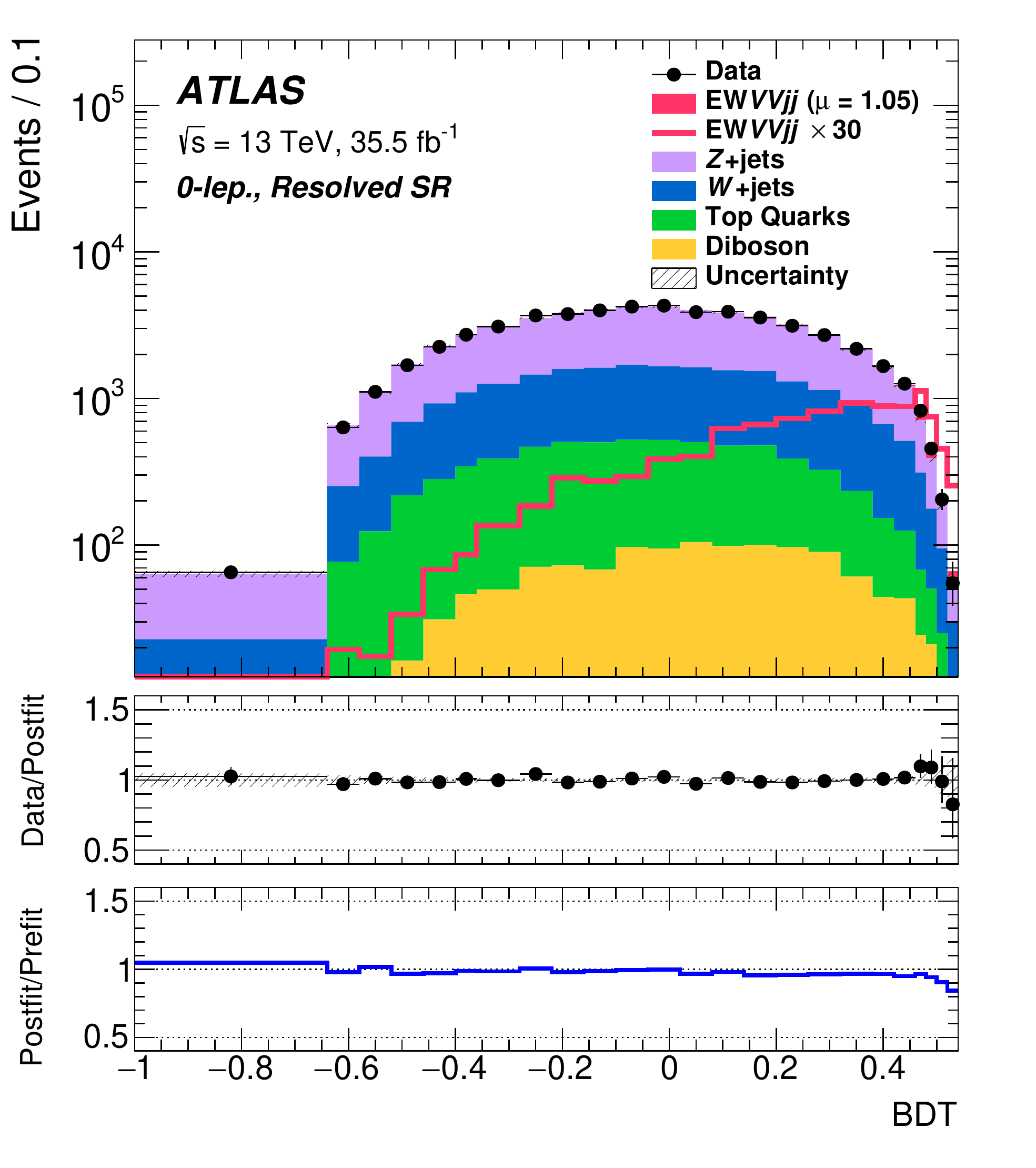}}
\caption{\label{fig:SR_BDT_0lep} Comparisons of the observed data and expected distributions of the BDT outputs
of the \zlep channel signal regions:  (a) high-purity and (b) low-purity merged signal regions; (c) the resolved signal region.
The background contributions after the global likelihood fit are shown as filled histograms.
The signal is shown as a filled histogram on top of
the fitted backgrounds normalized to the signal yield extracted from data ($\mu=\muVBSobsNom$),
and unstacked as an unfilled histogram, scaled by the factor indicated in the legend.
The entries in overflow are included in the last bin.
The middle pane shows the ratios of the observed data to the post-fit signal and background predictions.
The uncertainty in the total prediction, shown as bands, combines statistical and systematic contributions.
The bottom pane shows the ratios of the post-fit and pre-fit background predictions.
}
\end{figure}

\begin{figure}[!h!btp]
\centering
\subfigure[]{\includegraphics[width=0.45\textwidth]{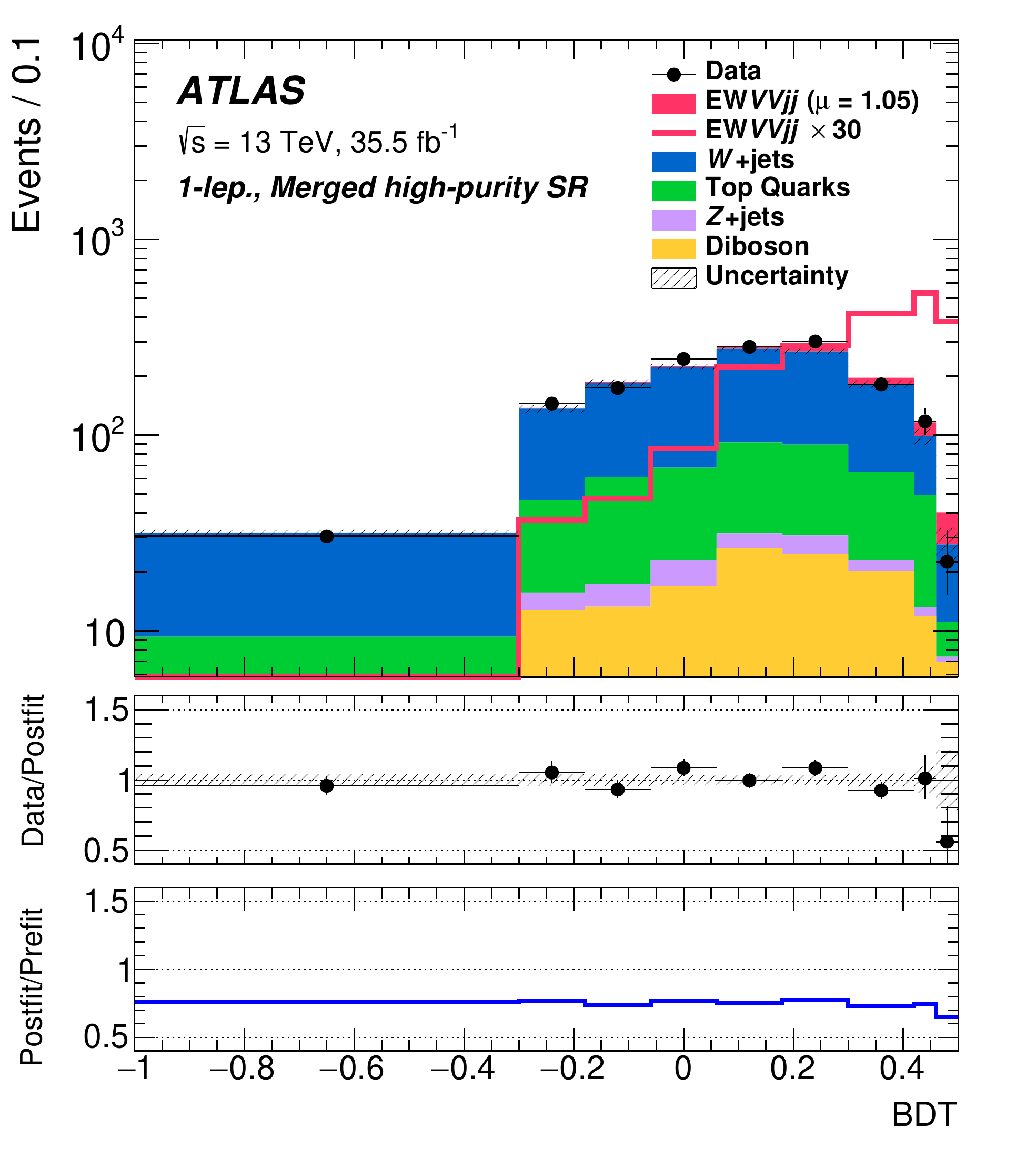}}
\subfigure[]{\includegraphics[width=0.45\textwidth]{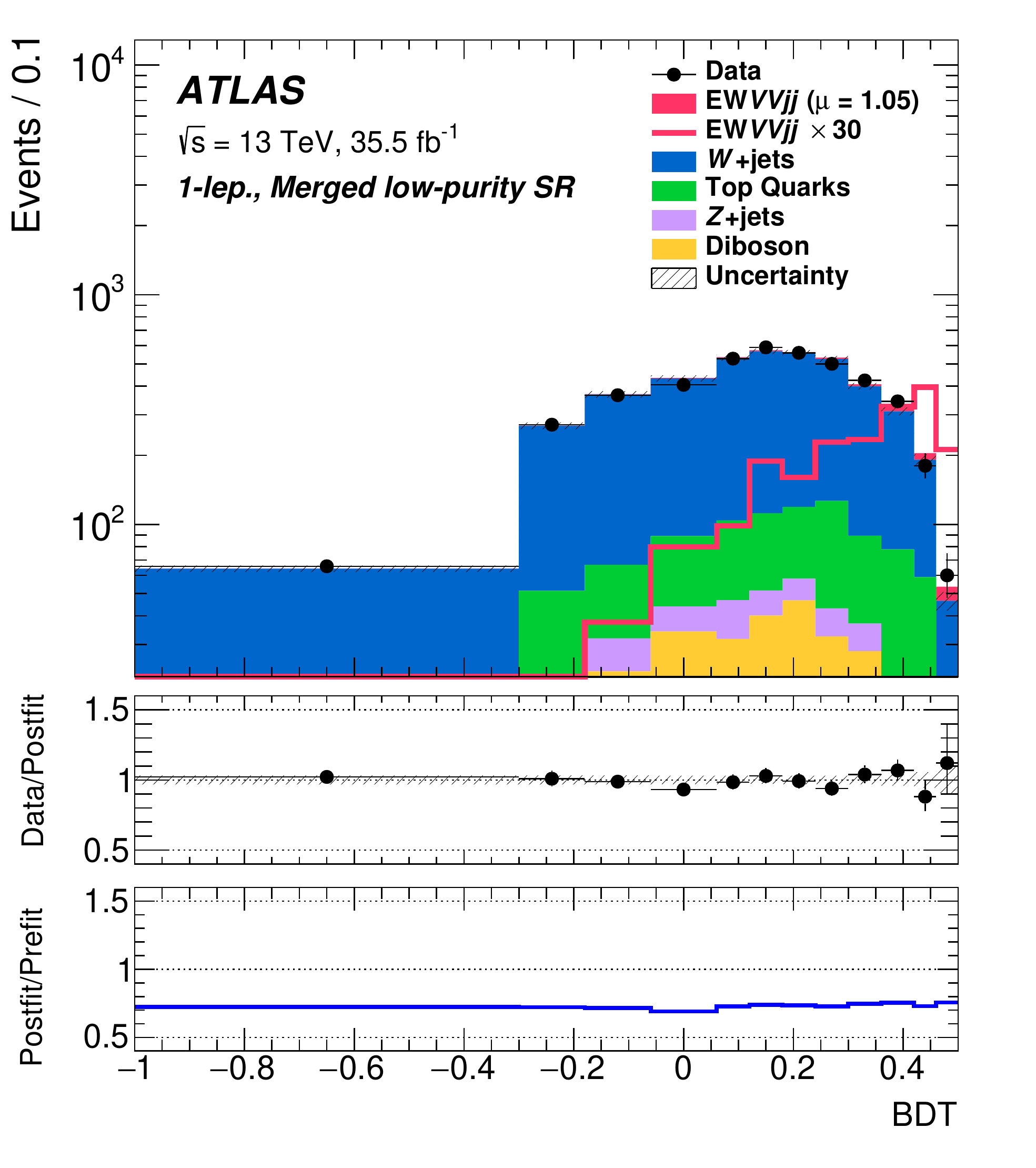}} \\
\subfigure[]{\includegraphics[width=0.45\textwidth]{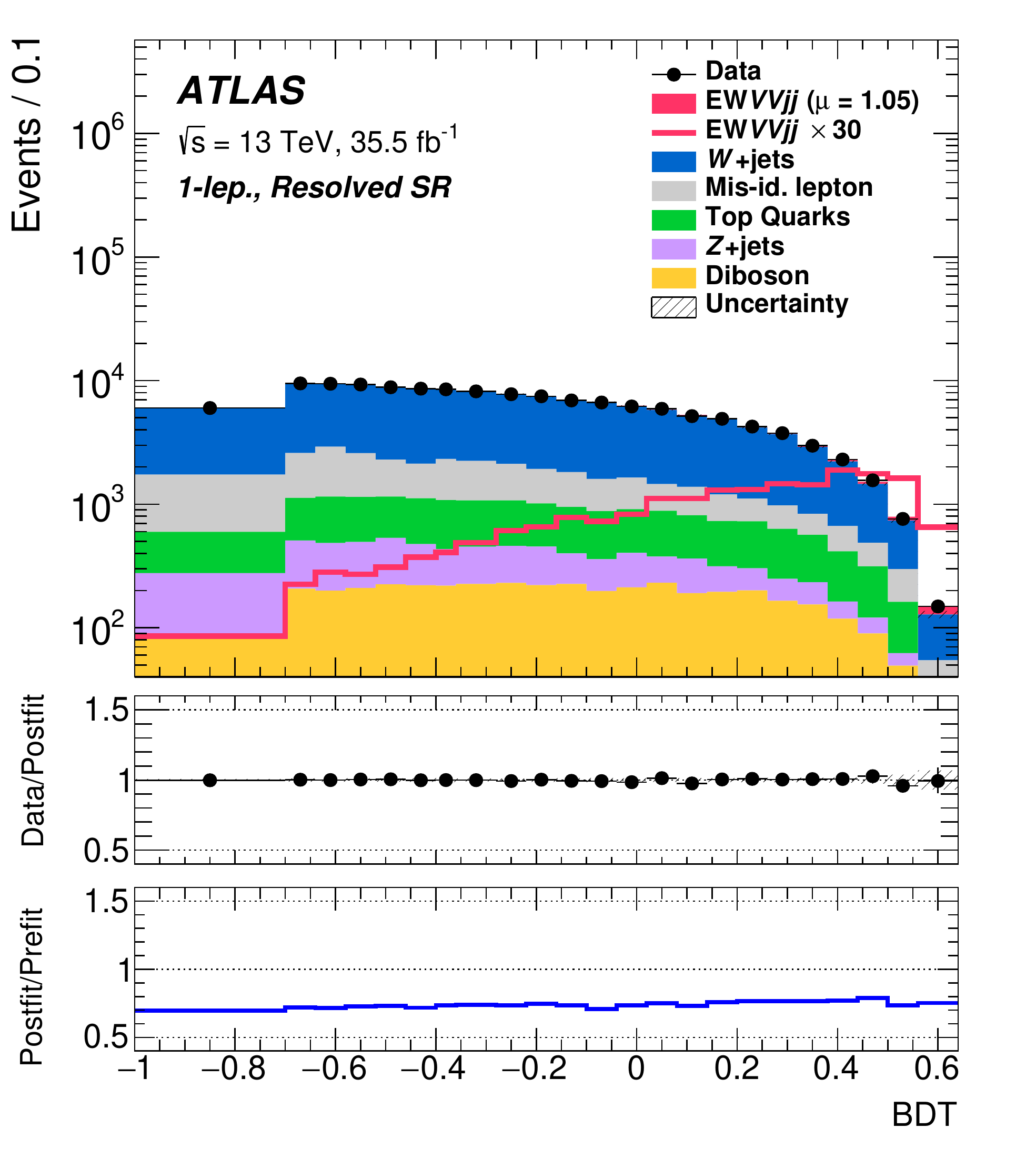}}
\caption{\label{fig:SR_BDT_1lep} Comparisons of the observed data and expected distributions of the BDT outputs
of the \olep channel signal regions:  (a) high-purity and (b) low-purity merged signal regions; (c) the resolved signal region.
The background contributions after the global likelihood fit are shown as filled histograms.
The signal is shown as a filled histogram on top of
the fitted backgrounds normalized to the signal yield extracted from data ($\mu=\muVBSobsNom$),
and unstacked as an unfilled histogram, scaled by the factor indicated in the legend.
The entries in overflow are included in the last bin.
The middle pane shows the ratios of the observed data to the post-fit signal and background predictions.
The uncertainty in the total prediction, shown as bands, combines statistical and systematic contributions.
The bottom pane shows the ratios of the post-fit and pre-fit background predictions.
}
\end{figure}

\begin{figure}[!h!btp]
\centering
\subfigure[]{\includegraphics[width=0.45\textwidth]{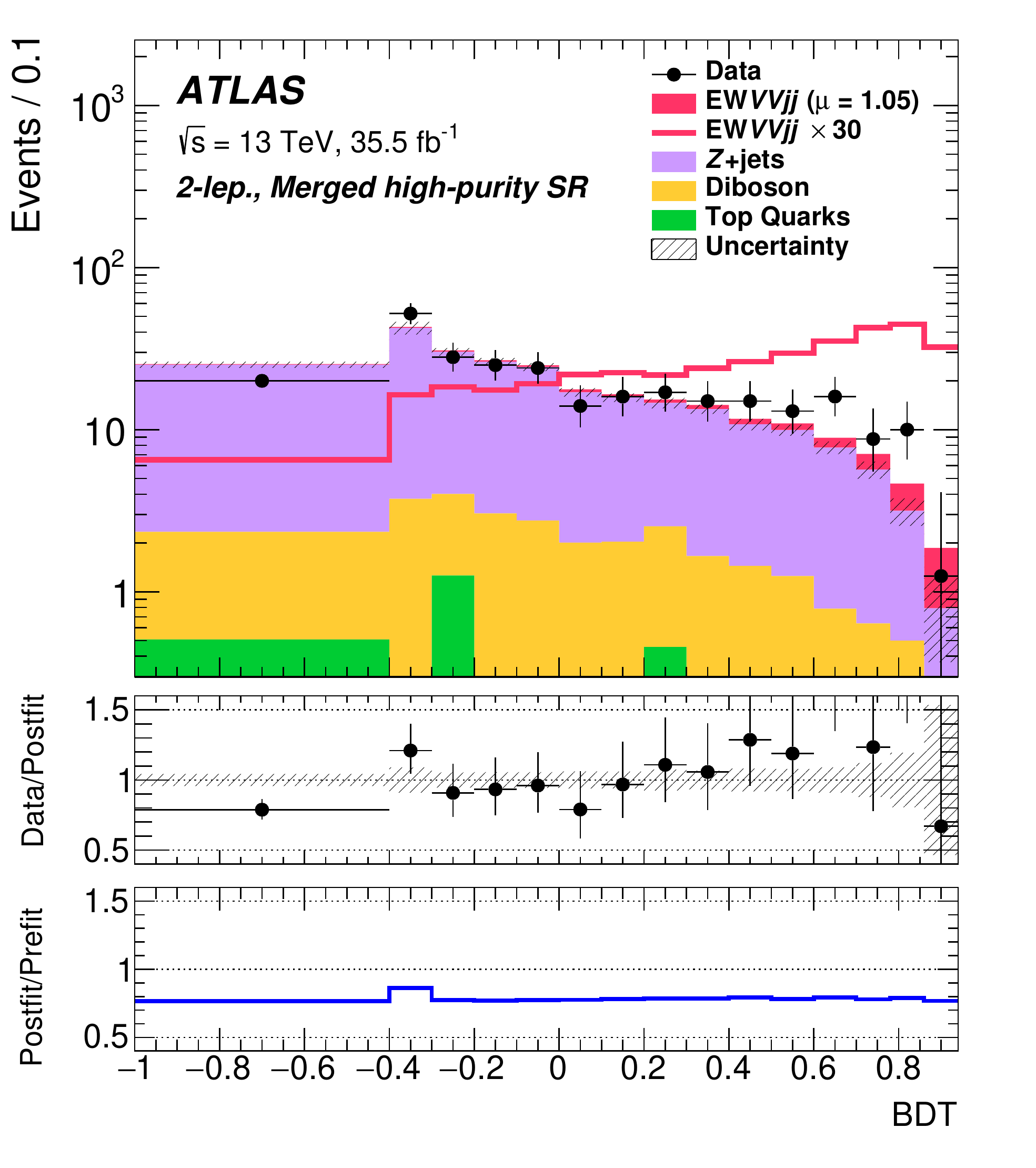}}
\subfigure[]{\includegraphics[width=0.45\textwidth]{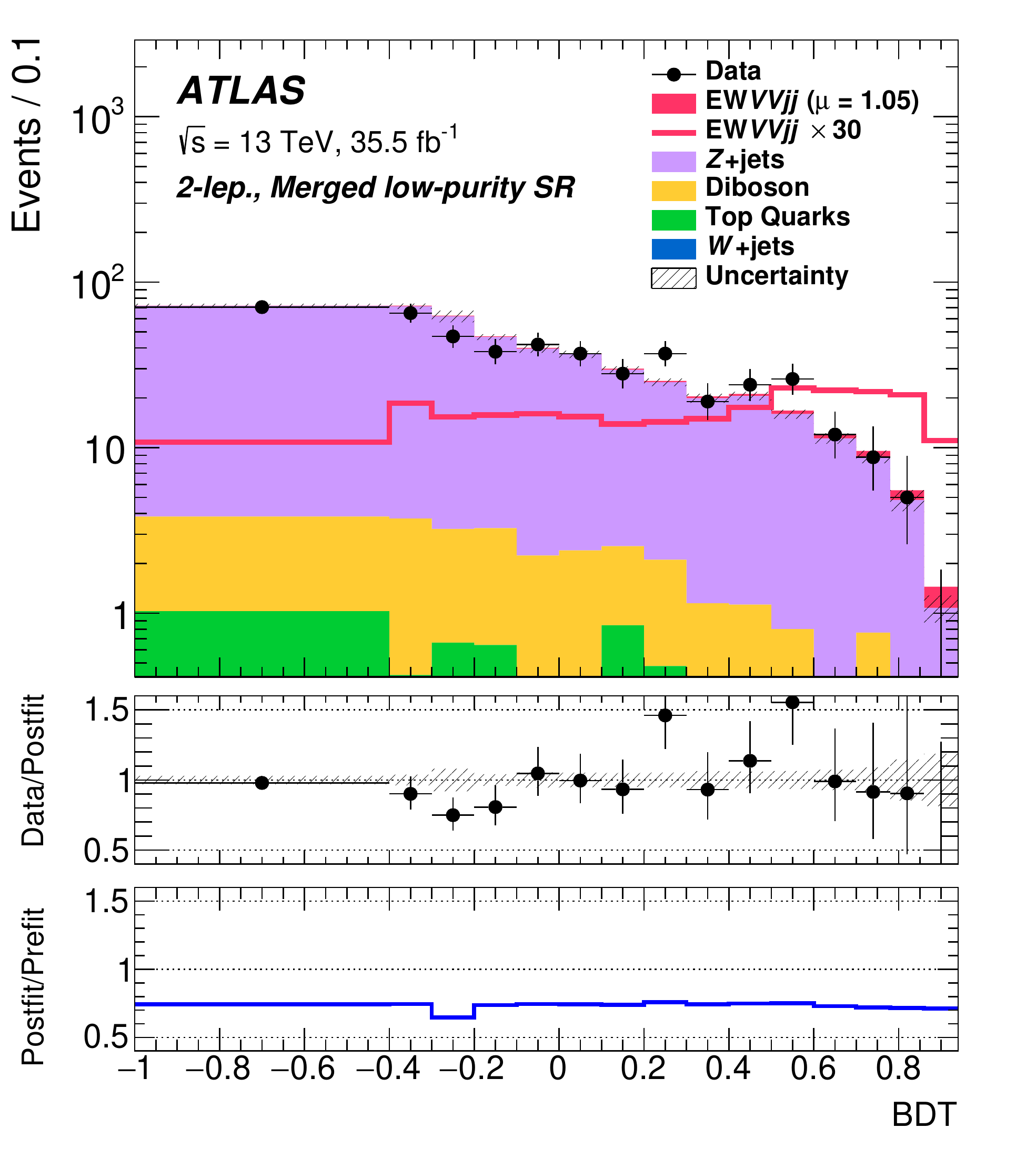}} \\
\subfigure[]{\includegraphics[width=0.45\textwidth]{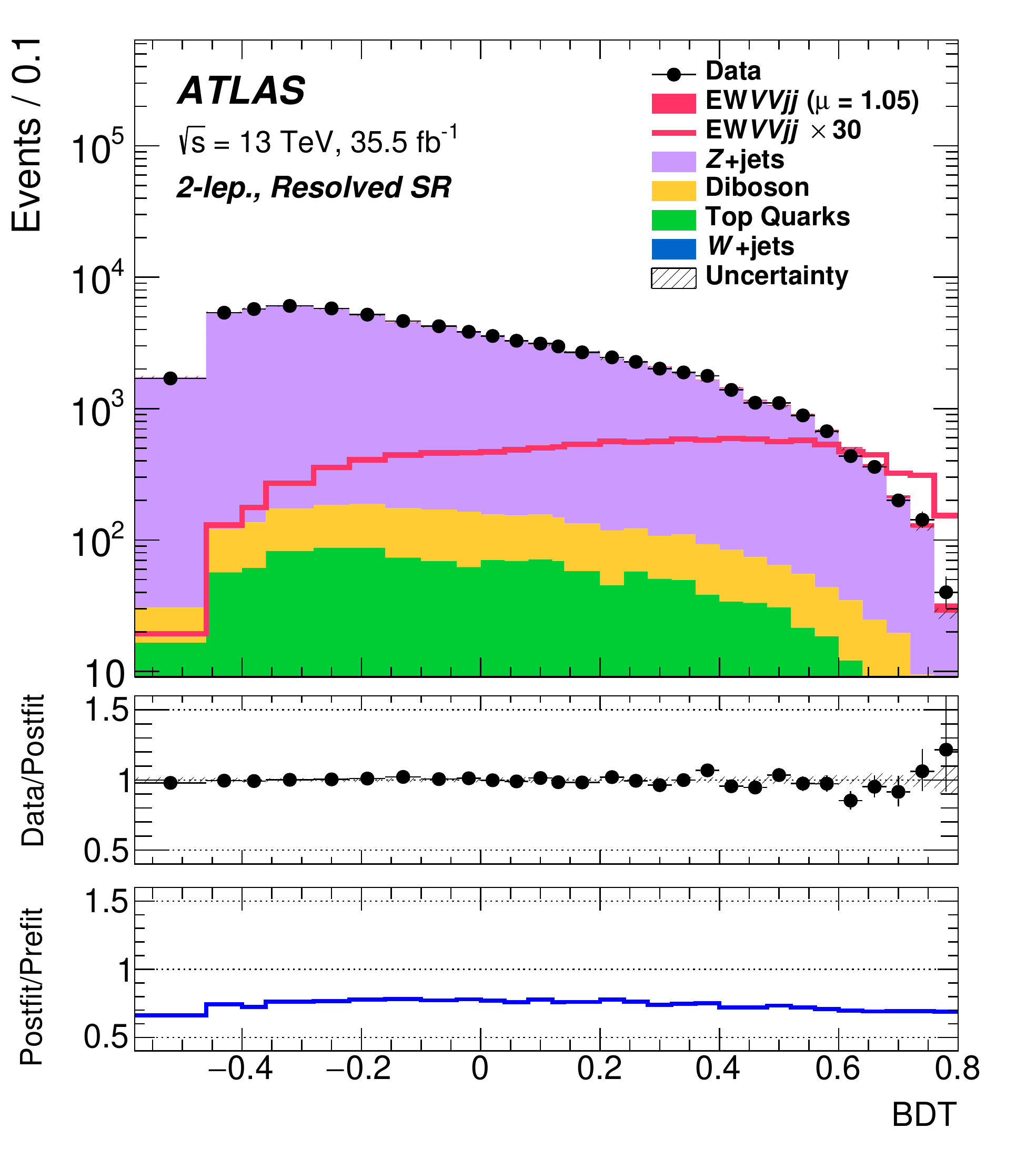}}
\caption{\label{fig:SR_BDT_2lep} Comparisons of the observed data and expected distributions of the BDT outputs
of the \tlep channel signal regions:  (a) high-purity and (b) low-purity merged signal regions; (c) the resolved signal region.
The background contributions after the global likelihood fit are shown as filled histograms.
The signal is shown as a filled histogram on top of
the fitted backgrounds normalized to the signal yield extracted from data ($\mu=\muVBSobsNom$),
and unstacked as an unfilled histogram, scaled by the factor indicated in the legend.
The entries in overflow are included in the last bin.
The middle pane shows the ratios of the observed data to the post-fit signal and background predictions.
The uncertainty in the total prediction, shown as bands, combines statistical and systematic contributions.
The bottom pane shows the ratios of the post-fit and pre-fit background predictions.
}
\end{figure}

The observed distributions of the BDT outputs in SRs used in the global likelihood fit are compared with the predictions,
shown in Figure~\ref{fig:SR_BDT_0lep} for the \zlep channel,
Figure~\ref{fig:SR_BDT_1lep} for the \olep channel, and Figure~\ref{fig:SR_BDT_2lep} for the \tlep channel.
The data distributions are reasonably well reproduced by the predicted contributions in all cases,
with the smallest $p$-value of 0.16 from the $\chi^{2}$ test~\cite{chitest} being for the \mjjtag\ distribution in the merged high-purity ZCR.
The numbers of events observed and estimated in the SRs are summarized in
Table~\ref{tab:vvqq_yieds_sbfit_SR} for the \zlep channel,
Table~\ref{tab:lvqq_yieds_sbfit_SR} for the \olep channel, and
Table~\ref{tab:llqq_yieds_sbfit_SR} for the \tlep channel.
The fitted value of the signal strength is
\begin{equation*}
\muVBSobs = \muVBSobsNom^{+\muVBSobsErrUp}_{-\muVBSobsErrDn} \
= \muVBSobsNom \pm \muVBSobsStat (\mathrm{stat.}) ^{+\muVBSobsSystUp}_{-\muVBSobsSystDn} (\mathrm{syst.}).
\label{eqn:muVBSobs}
\end{equation*}
The background-only hypothesis is excluded in data with a significance of \sigfVBSobs\ standard deviations,
compared with \sigfVBSexp\ standard deviations expected.
 
\begin{figure}[!h]
\begin{centering}
\includegraphics[width=0.65\textwidth]{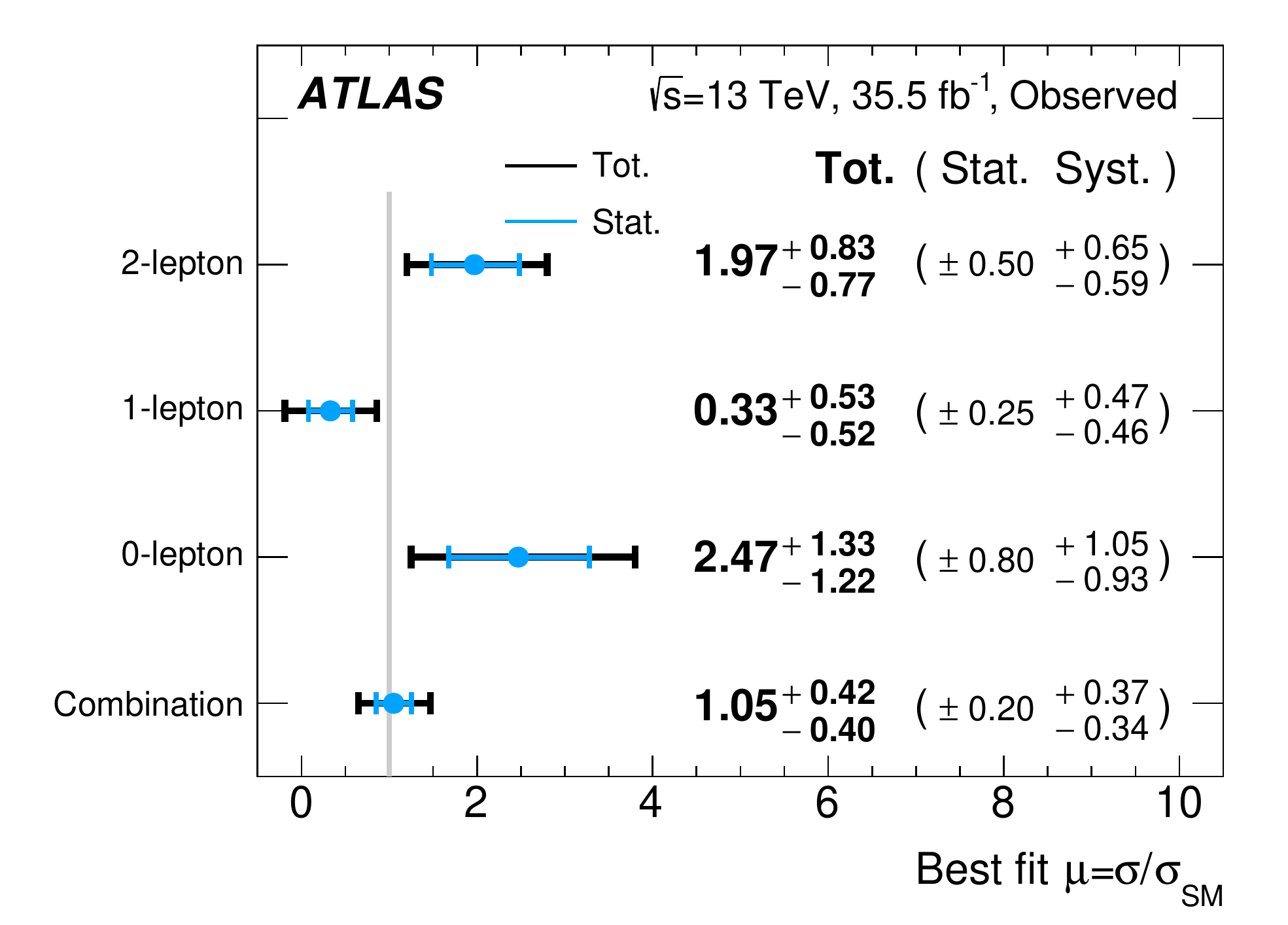}
\caption{
The fitted values of the signal-strength parameter \muVBSobs\ for the 0-, 1- and 2-lepton channels and their combination.
The individual \muVBSobs\ values for the lepton channels are obtained from a simultaneous fit with the signal-strength parameter
for each of the lepton channels floating independently.
The probability that the signal strengths measured in the three lepton channels are compatible is \pvalVBSobs.
}
\label{fig:fit_mu_VBS}
\end{centering}
\end{figure}

Figure~\ref{fig:fit_mu_VBS} shows the measured signal strength from the combined fit with a single signal-strength
fit parameter, and from a fit where each lepton channel has its own signal-strength parameter. The probability that the signal strengths measured in the three lepton channels are compatible is \pvalVBSobs.

% The next lines are included from the .//sections/yield_0lep_postfit_SR.tex input file

\begin{table}[!hbt]
\caption{\label{tab:vvqq_yieds_sbfit_SR} Numbers of events observed and predicted for signal and background processes in the
0-lepton channel signal regions,
obtained from signal-plus-background fits to the signal and control regions~(Section~\ref{sec:results}).
The signal yields are calculated after the fit with the observed signal strength of \muVBSobsNom\ applied.
The uncertainties combine statistical and systematic contributions.
The fit constrains the background estimate towards the observed data, which reduces the total background uncertainty
by correlating those uncertainties from the individual backgrounds.}
\centering
\vspace*{0.2cm}
\footnotesize

\begin{tabular}{ll
r@{\,\( \pm \)\,}l r@{\,\( \pm \)\,}l r@{\,\( \pm \)\,}l}
\toprule
 
\multicolumn{2}{c}{Sample} & \multicolumn{2}{c}{Resolved} & \multicolumn{2}{c}{Merged HP} & \multicolumn{2}{c}{Merged LP} \\
\hline
\multirow{5}{*}{Background}
& $W$ + jets         & $9200$ & $1300$& $259$ & $27$& $582$ & $56$ \\
& $Z$ + jets         & $19\,000$ & $1400$& $383$ & $29$& $955$ & $69$ \\
& Top quarks& $3280$ & $480$& $277$ & $28$& $276$ & $32$ \\
& Diboson   & $720$ & $120$& $69$ & $12$& $68$ & $14$ \\
\cmidrule(lr){3-8}
& Total     & $32\,100$ & $2000$& $988$ & $50$& $1881$ & $96$ \\
\midrule
\multirow{5}{*}{Signal}
& $W(\ell\nu)W(qq')$& $56$ & $22$& $8.0$ & $3.2$& $5.4$ & $2.2$ \\
& $W(\ell\nu)Z(qq) $& $12.0$ & $4.7$& $2.1$ & $0.8$& $1.6$ & $0.6$ \\
& $Z(\nu\nu)W(qq')$& $66$ & $25$& $9.0$ & $3.5$& $7.4$ & $2.9$ \\
& $Z(\nu\nu)Z(qq) $& $27$ & $10$& $5.1$ & $2.0$& $3.1$ & $1.2$ \\
\cmidrule(lr){3-8}
& Total     & $161$ & $35$& $24.3$ & $5.2$& $17.5$ & $3.9$ \\
\midrule
\multicolumn{2}{c}{SM}& $32\,300$ & $2000$& $1012$ & $50$& $1898$ & $96$ \\
\midrule
\multicolumn{2}{c}{Data} & \multicolumn{2}{c}{32\,299} & \multicolumn{2}{c}{1002} & \multicolumn{2}{c}{1935} \\
 
\bottomrule

\end{tabular}
\end{table}

% End of text imported from the .//sections/yield_0lep_postfit_SR.tex input file
% The next lines are included from the .//sections/yield_1lep_postfit_SR.tex input file

\begin{table}[!hbt]
\caption{\label{tab:lvqq_yieds_sbfit_SR} Numbers of events observed and predicted for signal and background processes in the
1-lepton channel signal regions,
obtained from signal-plus-background fits to the signal and control regions~(Section~\ref{sec:results}).
The signal yields are calculated after the fit with the observed signal strength of \muVBSobsNom\ applied.
The uncertainties combine statistical and systematic contributions.
The fit constrains the background estimate towards the observed data, which reduces the total background uncertainty
by correlating those uncertainties from the individual backgrounds.}
\centering
\vspace*{0.2cm}
\footnotesize

\begin{tabular}{ll
r@{\,\( \pm \)\,}l r@{\,\( \pm \)\,}l r@{\,\( \pm \)\,}l}
\toprule
 
\multicolumn{2}{c}{Sample} & \multicolumn{2}{c}{Resolved} & \multicolumn{2}{c}{Merged HP} & \multicolumn{2}{c}{Merged LP} \\
\hline
\multirow{6}{*}{Background}
& $W$ + jets         & $69\,100$ & $1900$& $1201$ & $65$& $2828$ & $97$ \\
& $Z$ + jets         & $2770$ & $370$& $39$ & $3$& $83$ & $6$ \\
& Top quarks& $7100$ & $1100$& $394$ & $56$& $422$ & $63$ \\
& Diboson   & $2660$ & $600$& $163$ & $35$& $229$ & $57$ \\
& Multijet  & $13\,400$ & $1600$ & \multicolumn{2}{c}{--}  & \multicolumn{2}{c}{--}  \\
\cmidrule(lr){3-8}
& Total     & $95\,100$ & $2800$& $1797$ & $93$& $3560$ & $130$ \\
\midrule
\multirow{6}{*}{Signal}
& $W(\ell\nu)W(qq')$& $330$ & $120$& $45$ & $17$& $34$ & $13$ \\
& $W(\ell\nu)Z(qq) $& $78$ & $29$& $11$ & $4$& $5$ & $2$ \\
\cmidrule(lr){3-8}
& Total     & $410$ & $130$& $57$ & $18$& $39$ & $13$ \\
\midrule
\multicolumn{2}{c}{SM}& $95\,500$ & $2800$& $1854$ & $95$& $3600$ & $130$ \\
\midrule
\multicolumn{2}{c}{Data} & \multicolumn{2}{c}{95\,366} & \multicolumn{2}{c}{1864} & \multicolumn{2}{c}{3571} \\
 
\bottomrule

\end{tabular}
\end{table}

% End of text imported from the .//sections/yield_1lep_postfit_SR.tex input file
% The next lines are included from the .//sections/yield_2lep_postfit_SR.tex input file

\begin{table}[!hbt]
\caption{\label{tab:llqq_yieds_sbfit_SR} Numbers of events observed and predicted for signal and background processes in the
2-lepton channel signal regions,
obtained from signal-plus-background fits to the signal and control regions~(Section~\ref{sec:results}).
The signal yields are calculated after the fit with the observed signal strength of \muVBSobsNom\ applied.
The uncertainties combine statistical and systematic contributions.
The fit constrains the background estimate towards the observed data, which reduces the total background uncertainty
by correlating those uncertainties from the individual backgrounds.}
\centering
\vspace*{0.2cm}
\footnotesize

\begin{tabular}{ll
r@{\,\( \pm \)\,}l r@{\,\( \pm \)\,}l r@{\,\( \pm \)\,}l}
\toprule
 
\multicolumn{2}{c}{Sample} & \multicolumn{2}{c}{Resolved} & \multicolumn{2}{c}{Merged HP} & \multicolumn{2}{c}{Merged LP} \\
\hline
\multirow{5}{*}{Background}
& $Z$ + jets         & $37\,090$ & $310$& $331$ & $14$& $775$ & $24$ \\
& Top quarks& $645$ & $99$& $5.8$ & $0.9$& $9.9$ & $2.7$ \\
& Diboson   & $830$ & $170$& $34.6$ & $7.6$& $36.7$ & $8.2$ \\
\cmidrule(lr){3-8}
& Total     & $38\,570$ & $370$& $371$ & $16$& $821$ & $25$ \\
\midrule
\multirow{4}{*}{Signal}
& $Z(\ell\ell)W(qq')$& $138$ & $53$& $8.6$ & $3.3$& $7.0$ & $2.7$ \\
& $Z(\ell\ell)Z(qq) $& $46$ & $18$& $4.3$ & $1.7$& $2.9$ & $1.1$ \\
\cmidrule(lr){3-8}
& Total     & $185$ & $56$& $12.9$ & $3.7$& $9.8$ & $2.9$ \\
\midrule
\multicolumn{2}{c}{SM}& $38\,760$ & $370$& $384$ & $17$& $831$ & $25$ \\
\midrule
\multicolumn{2}{c}{Data} & \multicolumn{2}{c}{38\,734} & \multicolumn{2}{c}{371} & \multicolumn{2}{c}{810} \\
 
\bottomrule

\end{tabular}
\end{table}

% End of text imported from the .//sections/yield_2lep_postfit_SR.tex input file
 
\clearpage
After the global maximum-likelihood fit, the uncertainties described in Section~\ref{sec:syst} are much reduced.
The effects of systematic uncertainties on the measurement after the fit are studied using
the signal-strength parameter \muVBSobs.
The relative uncertainties in the best-fit \muVBSobs\ value from the leading sources of systematic uncertainty
are shown in Table~\ref{tab:np_ranking_muobs}.
The individual sources of systematic uncertainty detailed in Section~\ref{sec:syst} are combined into categories.
Apart from the statistics of the data, the uncertainties with the largest impact on the sensitivity
of EW \VVjj\ production are from the modeling of background (\Zjets, \Wjets and QCD-induced diboson processes),
the modeling of the signal, $b$-tagging, and reconstruction of small-$R$ and large-$R$ jets.
 
\begin{table}[!ht]
\caption{\label{tab:np_ranking_muobs}
The symmetrized uncertainty $\sigma_{\mu}$ from each source in the best-fit signal-strength parameter \muVBSobs.
The floating normalizations include uncertainties of normalization scale factors for $Z$+jets, $W$+jets and
top quark contributions.
}
\centering
\begin{tabular}{lc}
\toprule
Uncertainty source    & $\sigma_{\mu}$ \\  \midrule
Total uncertainty                & 0.41   \\
Statistical                      & 0.20   \\
Systematic                       & 0.35   \\
\midrule
\multicolumn{2}{c}{Theoretical and modeling uncertainties}\\
\midrule
Floating normalizations          & 0.09   \\
\Zjets                           & 0.13   \\
\Wjets                           & 0.09   \\
\ttbar                           & 0.06   \\
Diboson                          & 0.09   \\
Multijet                         & 0.04   \\
Signal                           & 0.07   \\
MC statistics                    & 0.17   \\
\midrule
\multicolumn{2}{c}{Experimental uncertainties}\\
\midrule
Large-$R$ jets                   & 0.08   \\
Small-$R$ jets                   & 0.06   \\
Leptons                          & 0.02   \\
\MET                             & 0.04   \\
$b$-tagging                      & 0.07   \\
Pileup                          & 0.04   \\
Luminosity                       & 0.03   \\
 
\bottomrule
\end{tabular}
\end{table}

\clearpage
\subsection{Cross-section measurements}
\label{sec:res_fidxs}
 
The determination of the fiducial cross section is performed by scaling the measured signal strengths with the corresponding
SM predicted fiducial cross sections,
$\fidVBSobs =  \muVBSobs \cdot \fidVBSexp$.
It is assumed that there is no new physics that could cause sizable kinematic modifications of the background and signal.
Therefore, the only new physics signals that can be detected in an unbiased way are those leading to an enhanced EW \VVjj\
signal strength in the search region of this analysis.
The fiducial cross sections for EW \VVjj\ are measured in the merged and resolved fiducial phase-space regions described
in Section~\ref{sec:FidDef} and inclusively.
The merged HP SR and LP SR are combined to form one single merged fiducial phase-space region.
The systematic uncertainties of the measured fiducial cross sections include contributions from
experimental systematic uncertainties, theory modeling uncertainties in the backgrounds,
theory modeling uncertainties in the shapes of signal kinematic distributions,
and luminosity uncertainties.
The measured and SM predicted fiducial cross sections for EW \VVjj\ processes are summarized in Table~\ref{tab:obs_mu_fidxs},
where the measured values are obtained from two different simultaneous fits.
In the first fit, two signal-strength parameters are used, one for the merged category (both HP and LP), and the other one for
the resolved category; while in the second fit, a single signal-strength parameter is used.
The measured and SM predicted fiducial cross sections in each lepton channel are also reported in Table~\ref{tab:obs_mu_fidxs_ch}.
The measured values are obtained from a simultaneous fit where each lepton channel has its own signal-strength parameter,
and in each lepton channel the same signal-strength parameter is applied to both the merged and resolved categories.
The predictions are from \MGMCatNLO~2.4.3 at LO only, and no higher order corrections are included;
the theoretical uncertainties due to the PDF, missing higher-order corrections, and parton-shower modeling are estimated
as described in Section~\ref{sec:syst}.
The measured fiducial cross sections are generally consistent with the SM predictions.

\begin{table}[htb]
\centering
\caption{Summary of predicted and measured fiducial cross sections for EW \VVjj\ production.
The three lepton channels are combined.
For the measured fiducial cross sections in the merged and resolved categories, two signal-strength parameters are used in the combined fit,
one for the merged category and the other one for the resolved category;
while for the measured fiducial cross section in the inclusive fiducial phase space,
a single signal-strength parameter is used.
For the SM predicted cross section, the error is the theoretical uncertainty~(theo.).
For the measured cross section, the first error is the statistical uncertainty~(stat.), and the second error is the
systematic uncertainty~(syst.).
\label{tab:obs_mu_fidxs}
}
 
\begin{tabular}{l
r@{\,\( \pm \)\,}l r@{\,\( \pm \)\,}l@{\,}l}
\toprule
Fiducial phase space  & \multicolumn{2}{c}{Predicted \fidVBSexp\ [fb]} & \multicolumn{3}{c}{Measured \fidVBSobs\ [fb]} \\
\midrule
\\ [-3mm]
Merged    &  $\fidVBSBstexpNom$ & $\fidVBSBstexpSyst \,(\mathrm{theo.})$  & $\fidVBSBstobsNom$ & $\fidVBSBstobsStat \,(\mathrm{stat.})$ & $^{+\fidVBSBstobsSystUp}_{-\fidVBSBstobsSystDn} \,(\mathrm{syst.})$ \\
\\ [-3mm]
Resolved  &  $\fidVBSResexpNom$ & $\fidVBSResexpSyst \,(\mathrm{theo.})$  & $\fidVBSResobsNom$ & $\fidVBSResobsStat \,(\mathrm{stat.})$ & $^{+\fidVBSResobsSystUp}_{-\fidVBSResobsSystDn} \,(\mathrm{syst.})$ \\
\\ [-3mm]
\midrule
\\ [-3mm]
Inclusive &  $\fidVBSexpNom$ & $\fidVBSexpSyst \,(\mathrm{theo.})$  & $\fidVBSobsNom$ & $\fidVBSobsStat \,(\mathrm{stat.})$ & $^{+\fidVBSobsSystUp}_{-\fidVBSobsSystDn} \,(\mathrm{syst.})$ \\
\\ [-3mm]
\bottomrule
\end{tabular}

\end{table}

\begin{table}[htb]
\centering
\caption{Summary of predicted and measured fiducial cross sections for EW \VVjj\ production
in the three lepton channels.
The measured values are obtained from a simultaneous fit where each lepton channel has its own signal-strength parameter,
and in each lepton channel the same signal-strength parameter is applied to both the merged and resolved categories.
For the SM predicted cross section, the error is the theoretical uncertainty~(theo.).
For the measured cross section, the first error is the statistical uncertainty~(stat.), and the second error is the
systematic uncertainty~(syst.).
\label{tab:obs_mu_fidxs_ch}
}
 
\begin{tabular}{ll
r@{\,\( \pm \)\,}l r@{\,\( \pm \)\,}l@{\,}l}
\toprule
 
\multicolumn{2}{c}{Fiducial phase space}  & \multicolumn{2}{c}{Predicted \fidVBSexp\ [fb]} & \multicolumn{3}{c}{Measured \fidVBSobs\ [fb]} \\
\midrule
 
\multirow{4}{*}{Merged}
& \zlep &  $\fidVBSZlepBstexpNom $ & $ \fidVBSZlepBstexpSyst \,(\mathrm{theo.})$  & $\fidVBSZlepBstobsNom $ & $ \fidVBSZlepBstobsStat \,(\mathrm{stat.}) $ & $^{+\fidVBSZlepBstobsSystUp}_{-\fidVBSZlepBstobsSystDn} \,(\mathrm{syst.})$ \\
\\ [-3mm]
& \olep &  $\fidVBSOlepBstexpNom $ & $ \fidVBSOlepBstexpSyst \,(\mathrm{theo.})$  & $\fidVBSOlepBstobsNom $ & $ \fidVBSOlepBstobsStat \,(\mathrm{stat.}) $ & $^{+\fidVBSOlepBstobsSystUp}_{-\fidVBSOlepBstobsSystDn} \,(\mathrm{syst.})$ \\
\\ [-3mm]
& \tlep &  $\fidVBSTlepBstexpNom $ & $ \fidVBSTlepBstexpSyst \,(\mathrm{theo.})$  & $\fidVBSTlepBstobsNom $ & $ \fidVBSTlepBstobsStat \,(\mathrm{stat.}) $ & $^{+\fidVBSTlepBstobsSystUp}_{-\fidVBSTlepBstobsSystDn} \,(\mathrm{syst.})$ \\
\\ [-3mm]
 
\midrule
\multirow{4}{*}{Resolved}
& \zlep &  $\fidVBSZlepResexpNom $ & $ \fidVBSZlepResexpSyst \,(\mathrm{theo.})$  & $\fidVBSZlepResobsNom $ & $ \fidVBSZlepResobsStat \,(\mathrm{stat.}) $ & $^{+\fidVBSZlepResobsSystUp}_{-\fidVBSZlepResobsSystDn} \,(\mathrm{syst.})$ \\
\\ [-3mm]
& \olep &  $\fidVBSOlepResexpNom $ & $ \fidVBSOlepResexpSyst \,(\mathrm{theo.})$  & $\fidVBSOlepResobsNom $ & $ \fidVBSOlepResobsStat \,(\mathrm{stat.}) $ & $^{+\fidVBSOlepResobsSystUp}_{-\fidVBSOlepResobsSystDn} \,(\mathrm{syst.})$ \\
\\ [-3mm]
& \tlep &  $\fidVBSTlepResexpNom $ & $ \fidVBSTlepResexpSyst \,(\mathrm{theo.})$  & $\fidVBSTlepResobsNom $ & $ \fidVBSTlepResobsStat \,(\mathrm{stat.}) $ & $^{+\fidVBSTlepResobsSystUp}_{-\fidVBSTlepResobsSystDn} \,(\mathrm{syst.})$ \\
\\ [-3mm]
 
\midrule
\multirow{4}{*}{Inclusive}
& \zlep &  $\fidVBSZlepexpNom $ & $ \fidVBSZlepexpSyst \,(\mathrm{theo.})$  & $\fidVBSZlepobsNom $ & $ \fidVBSZlepobsStat \,(\mathrm{stat.}) $ & $^{+\fidVBSZlepobsSystUp}_{-\fidVBSZlepobsSystDn} \,(\mathrm{syst.})$ \\
\\ [-3mm]
& \olep &  $\fidVBSOlepexpNom $ & $ \fidVBSOlepexpSyst \,(\mathrm{theo.})$  & $\fidVBSOlepobsNom $ & $ \fidVBSOlepobsStat \,(\mathrm{stat.}) $ & $^{+\fidVBSOlepobsSystUp}_{-\fidVBSOlepobsSystDn} \,(\mathrm{syst.})$ \\
\\ [-3mm]
& \tlep &  $\fidVBSTlepexpNom $ & $ \fidVBSTlepexpSyst \,(\mathrm{theo.})$  & $\fidVBSTlepobsNom $ & $ \fidVBSTlepobsStat \,(\mathrm{stat.}) $ & $^{+\fidVBSTlepobsSystUp}_{-\fidVBSTlepobsSystDn} \,(\mathrm{syst.})$ \\
\\ [-3mm]
 
\bottomrule
\end{tabular}

\end{table}

% End of text imported from the .//./sections/Results.tex input file
 
\FloatBarrier
\section{Conclusion}
\label{sec:conclusion}
% The next lines are included from the .//./sections/Conclusions.tex input file
A measurement of \VVjj~($V=W,Z$) electroweak production using \cms\ $pp$ collisions at the LHC is presented.
The data were collected with the ATLAS detector in 2015 and 2016 and correspond to a total integrated luminosity of \lumi.
The study explores the final states with one boson decaying leptonically, and the other boson decaying into a pair of quarks,
identified either as two separate jets or as one large-radius jet.
 
The \VVjj\ electroweak production cross section is measured with a significance of \sigfVBSobs\ standard deviations
over the background-only hypothesis. The expected significance is \sigfVBSexp\ standard deviations.
The measured signal strength relative to the leading-order SM prediction is
$\muVBSobs = \muVBSobsNom \pm \muVBSobsStat (\mathrm{stat.}) ^{+\muVBSobsSystUp}_{-\muVBSobsSystDn} (\mathrm{syst.})$.
The fiducial cross section of \VVjj\ electroweak production is measured to be
$\fidVBSobs = \fidVBSobsNom \pm \fidVBSobsStat (\mathrm{stat.}) ^{+\fidVBSobsSystUp}_{-\fidVBSobsSystDn} (\mathrm{syst.})$~fb.
% End of text imported from the .//./sections/Conclusions.tex input file

\section*{Acknowledgments}
% The next lines are included from the .//acknowledgements/Acknowledgements.tex input file
 
We thank CERN for the very successful operation of the LHC, as well as the
support staff from our institutions without whom ATLAS could not be
operated efficiently.
 
We acknowledge the support of ANPCyT, Argentina; YerPhI, Armenia; ARC, Australia; BMWFW and FWF, Austria; ANAS, Azerbaijan; SSTC, Belarus; CNPq and FAPESP, Brazil; NSERC, NRC and CFI, Canada; CERN; CONICYT, Chile; CAS, MOST and NSFC, China; COLCIENCIAS, Colombia; MSMT CR, MPO CR and VSC CR, Czech Republic; DNRF and DNSRC, Denmark; IN2P3-CNRS, CEA-DRF/IRFU, France; SRNSFG, Georgia; BMBF, HGF, and MPG, Germany; GSRT, Greece; RGC, Hong Kong SAR, China; ISF and Benoziyo Center, Israel; INFN, Italy; MEXT and JSPS, Japan; CNRST, Morocco; NWO, Netherlands; RCN, Norway; MNiSW and NCN, Poland; FCT, Portugal; MNE/IFA, Romania; MES of Russia and NRC KI, Russian Federation; JINR; MESTD, Serbia; MSSR, Slovakia; ARRS and MIZ\v{S}, Slovenia; DST/NRF, South Africa; MINECO, Spain; SRC and Wallenberg Foundation, Sweden; SERI, SNSF and Cantons of Bern and Geneva, Switzerland; MOST, Taiwan; TAEK, Turkey; STFC, United Kingdom; DOE and NSF, United States of America. In addition, individual groups and members have received support from BCKDF, CANARIE, CRC and Compute Canada, Canada; COST, ERC, ERDF, Horizon 2020, and Marie Sk{\l}odowska-Curie Actions, European Union; Investissements d' Avenir Labex and Idex, ANR, France; DFG and AvH Foundation, Germany; Herakleitos, Thales and Aristeia programmes co-financed by EU-ESF and the Greek NSRF, Greece; BSF-NSF and GIF, Israel; CERCA Programme Generalitat de Catalunya, Spain; The Royal Society and Leverhulme Trust, United Kingdom.
 
The crucial computing support from all WLCG partners is acknowledged gratefully, in particular from CERN, the ATLAS Tier-1 facilities at TRIUMF (Canada), NDGF (Denmark, Norway, Sweden), CC-IN2P3 (France), KIT/GridKA (Germany), INFN-CNAF (Italy), NL-T1 (Netherlands), PIC (Spain), ASGC (Taiwan), RAL (UK) and BNL (USA), the Tier-2 facilities worldwide and large non-WLCG resource providers. Major contributors of computing resources are listed in Ref.~\cite{ATL-GEN-PUB-2016-002}.
 
% End of text imported from the .//acknowledgements/Acknowledgements.tex input file

\printbibliography

\clearpage
% ATLAS Collaboration author list
% Reference date of STDM-2017-20 is 2019-01-14
% Author list last updated on date 24-JUN-19
% Data extracted on 24-Jun-2019 for paper reference STDM-2017-20
% at 3:16pm
 
\begin{flushleft}
{\Large The ATLAS Collaboration}

\bigskip

G.~Aad$^\textrm{\scriptsize 101}$,    
B.~Abbott$^\textrm{\scriptsize 128}$,    
D.C.~Abbott$^\textrm{\scriptsize 102}$,    
O.~Abdinov$^\textrm{\scriptsize 13,*}$,    
A.~Abed~Abud$^\textrm{\scriptsize 70a,70b}$,    
K.~Abeling$^\textrm{\scriptsize 53}$,    
D.K.~Abhayasinghe$^\textrm{\scriptsize 93}$,    
S.H.~Abidi$^\textrm{\scriptsize 167}$,    
O.S.~AbouZeid$^\textrm{\scriptsize 40}$,    
N.L.~Abraham$^\textrm{\scriptsize 156}$
% ...
\bigskip
\\

% ...
$^{13}$Institute of Physics, Azerbaijan Academy of Sciences, Baku; Azerbaijan.\\
% ...
$^{40}$Niels Bohr Institute, University of Copenhagen, Copenhagen; Denmark.\\
% ...
$^{53}$II. Physikalisches Institut, Georg-August-Universit\"{a}t G\"ottingen, G\"ottingen; Germany.\\
% ...
$^{70}$$^{(a)}$INFN Sezione di Pavia;$^{(b)}$Dipartimento di Fisica, Universit\`a di Pavia, Pavia; Italy.\\
% ...
$^{93}$Department of Physics, Royal Holloway University of London, Egham; United Kingdom.\\
% ...
$^{101}$CPPM, Aix-Marseille Universit\'e, CNRS/IN2P3, Marseille; France.\\
$^{102}$Department of Physics, University of Massachusetts, Amherst MA; United States of America.\\
% ...
$^{128}$Homer L. Dodge Department of Physics and Astronomy, University of Oklahoma, Norman OK; United States of America.\\
% ...
$^{156}$Department of Physics and Astronomy, University of Sussex, Brighton; United Kingdom.\\
% ...
$^{167}$Department of Physics, University of Toronto, Toronto ON; Canada.\\
% ...
$^{a}$ Also at Centre for High Performance Computing, CSIR Campus, Rosebank, Cape Town; South Africa.\\
$^{b}$ Also at CERN, Geneva; Switzerland.\\
% ...
$^{*}$ Deceased

\end{flushleft}

% Created with Glance <Atlas.Glance@cern.ch>
 
\end{document}